\documentclass[journal]{IEEEtran}  %
\usepackage{multicol}
\usepackage[pageanchor=true,plainpages=false, pdfpagelabels, bookmarks,bookmarksnumbered,hidelinks]{hyperref}
\usepackage{pdfsync}		%
\usepackage{amsmath}		%
\usepackage{amssymb}		%
\usepackage{bm}             %
\usepackage{graphicx}		%
\usepackage[footnotesize]{caption}
\usepackage[listofformat=subsimple,
labelformat=simple]{subfig}
\usepackage{psfrag}   	%
\usepackage{color}			%
\usepackage{balance}    %
\usepackage[sort]{cite}
\usepackage{CommonSource/mathnotation} %
\usepackage{booktabs}
\usepackage{multirow}
\usepackage[multiple]{footmisc}
\usepackage{enumitem}

\newcommand{\subparagraph}{}

\usepackage{marginnote}

\makeatletter
\renewcommand*\env@matrix[1][*\c@MaxMatrixCols c]{%
\hskip -\arraycolsep
\let\@ifnextchar\new@ifnextchar
\array{#1}}
\makeatother

\newcommand{\inertialcost}{E}

\title{
The Geometry of Optimal Gaits for\\ Inertia-dominated Kinematic Systems
}

\author{Ross L. Hatton, Zachary Brock, Shuoqi Chen, Howie Choset, Hossein Faraji, Ruijie Fu, \\ Nathan Justus and Suresh Ramasamy%
\thanks{}
\thanks{R. L. Hatton, Z. Brock, H. Faraji, N. Justus and S. Ramasamy are with the  Collaborative Robotics and Intelligent Systems (CoRIS) Institute at Oregon State University, Corvallis, OR USA. Ross.Hatton@oregonstate.edu}%
\thanks{S. Chen, H. Choset, and R. Fu are with the Robotics Institute at Carnegie Mellon University, Pittsburgh, PA USA.
}%
}

\begin{document}

\maketitle
\thispagestyle{empty}
\pagestyle{plain}

\begin{abstract}
Isolated mechanical systems---e.g., those floating in space, in free-fall, or on a frictionless surface---are able to achieve net rotation by cyclically changing their shape, even if they have no net angular momentum. Similarly, swimmers immersed in ``perfect fluids" are able to use cyclic shape changes to both translate and rotate even if the swimmer-fluid system has no net linear or angular momentum. Finally, systems fully constrained by direct nonholonomic constraints (e.g., passive wheels) can push against these constraints to move through the world. Previous work has demonstrated that the net displacement induced by these shape changes corresponds to the amount of \emph{constraint curvature} that the gaits enclose.

To properly assess or optimize the utility of a gait, however, we must also consider the time or resources required to execute it: A gait that produces a small displacement per cycle, but that can be executed in a short time, may produce a faster average velocity than a gait that produces a large displacement per cycle, but takes much longer to complete a cycle at the same average instantaneous effort.

In this paper, we consider two effort-based cost functions for assessing the costs associated with executing these cycles. For each of these cost functions, we demonstrate that fixing the average instantaneous cost to a unit value allows us to  transform the effort costs into  time-to-execute costs for any given gait cycle.  We then illustrate how the interaction between the constraint curvature and these costs leads to characteristic geometries for optimal cycles, in which the gait trajectories resemble elastic hoops distended from within by internal pressures.

\end{abstract}

\section{Introduction}

Isolated mechanical systems---e.g., those floating in space, in free-fall, or on a frictionless surface---are able to achieve net rotation by cyclically changing their shape, even if they have no net angular momentum~\cite{montgomery1993gauge,walsh95}. Similarly, swimmers immersed in ``perfect fluids" are able to use cyclic shape changes to both translate and rotate even if the swimmer-fluid system has no net linear or angular momentum~\cite{Kanso:2005}. The operating principle in both cases is that the systems' moments of inertia (and in the case of perfect-fluid swimming, the systems' fluid-added masses) depend on the system shape. By moving portions of the body forward in low-inertia-configurations and backward in high-inertia configurations, systems can generate net displacements even while their total momentum remains zero, i.e., variable inertia means that conservation of momentum does not lead to conservation of position.

\begin{figure}%
\centering
\includegraphics[width=.35\textwidth]{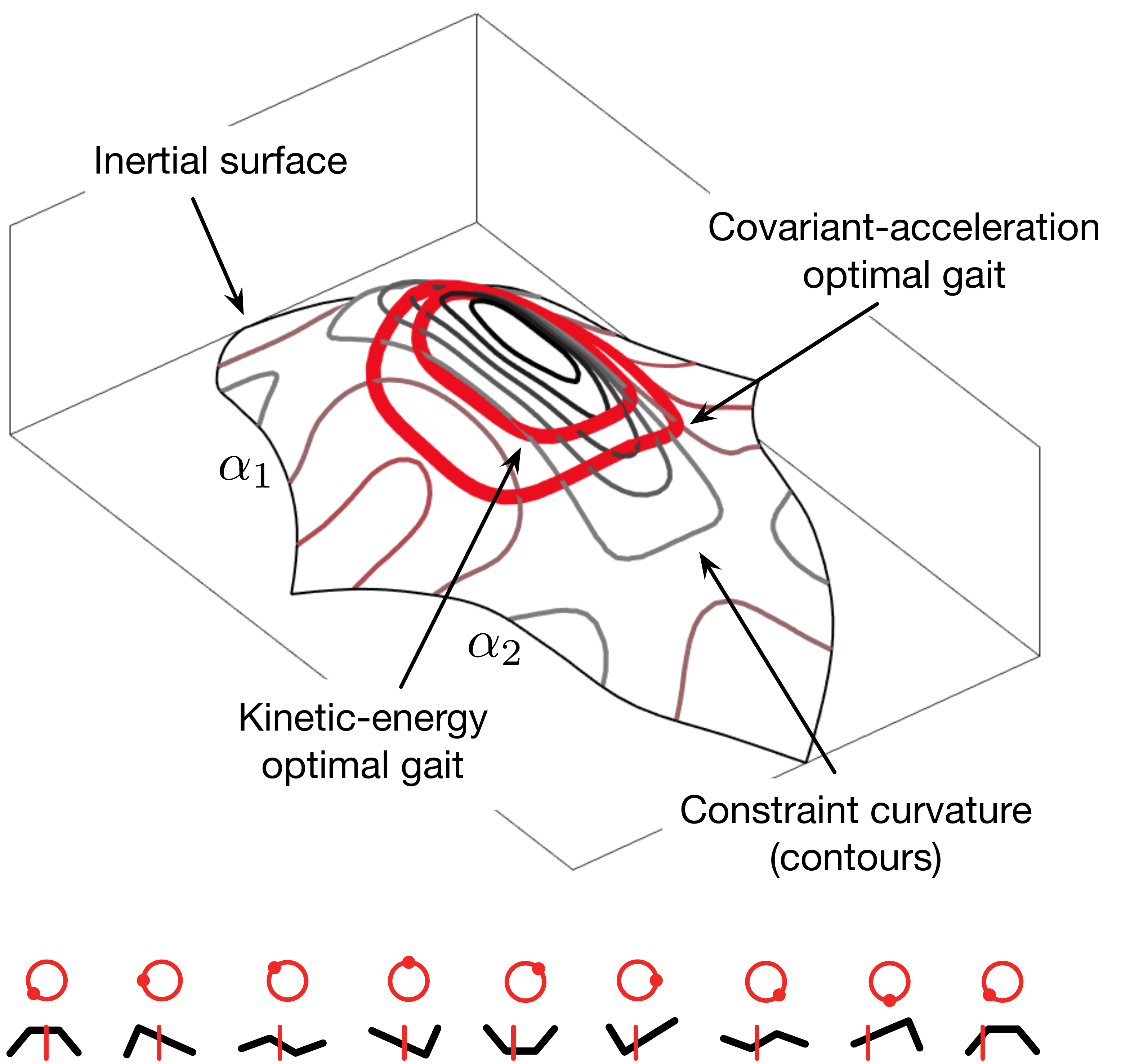}
\caption{The net position displacement of an inertial-kinematic system over a gait cycle corresponds to the amount of \emph{constraint curvature} that the gait's trajectory encloses on the system's shape space.  The cost of executing a gait can be measured in terms of the kinetic energy or acceleration required to follow the trajectory, which correspond to the length and roundness of the trajectory over a surface shaped by the system's inertial distribution. Gaits that maximize speed for a given energy or acceleration budget balance these two influences, with the constraint curvature acting as an ``inflationary pressure" favoring large-amplitude cycles, and the pathlength and roundness terms providing a ``surface tension" and ``bending stiffness" that constrain the sizes and aspect ratios of the optimal gait cycles. As a general rule, the acceleration-optimal gait for a system will be both larger and rounder than the energy-optimal gait.
}
\label{fig:InertialIntroFig}
\end{figure}

Given this principle, it is natural to ask ``What shape trajectories best exploit the changes in inertia to generate system motion?" Answering this question involves investigating two subquestions,
\begin{enumerate}
\item How much displacement is achieved over a given cycle?
\item How much time or energy does it cost to execute a given cycle?
\end{enumerate}
and then dividing the displacement by the cost to get an overall efficiency (in terms of opportunity or resources) for the motion.

In previous work, e.g.,~\cite{walsh95,Melli:2006,Hatton:2013TRO:Swimming,9158895}, we and others have addressed the first question, demonstrating that the displacement of an isolated or perfect-fluid system induced by a gait cycle corresponds to the amount of \emph{constraint curvature} in the system dynamics that the gait trajectory encloses: gaits that produce large displacements enclose strongly sign-definite regions of this curvature, and gaits that produce zero displacement enclose sign-balanced regions. As yet, however, we believe that geometric characterizations of the cost of executing different gaits under inertial dynamics (and therefore the optimality of such motions) have yet to be explored.

\begin{figure*}%
\centering
\includegraphics[width=.85\textwidth]{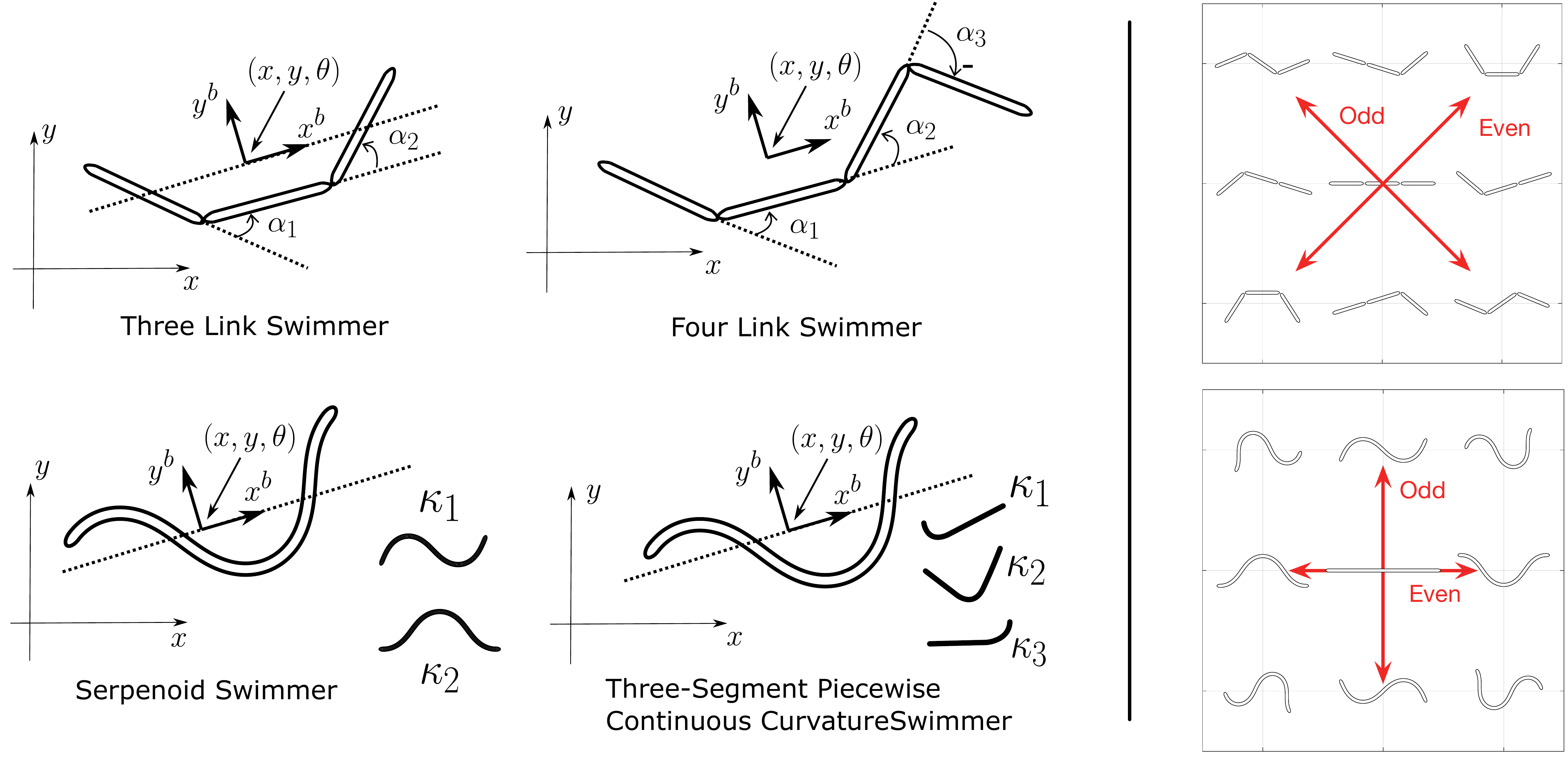}
\caption{Geometry and configuration variables of some of the example systems. The systems in the top row are articulated swimmers, while the bottom row consists of continuous curvature swimmers. The shape of the systems in the first column are described by two shape variables whereas the shape of the systems in the second column are described by three shape variables. Articulated and continuous shape spaces can both be described as having ``even" and ``odd" axes, (in which the joint angles or body curvature have even or odd symmetry about the system's midpoint). These axes are illustrated in the rightmost column for systems with two shape variables; adding midpoint bending as a third shape variable creates a second ``even" axis, and adding a fourth mode provides a second ``odd" mode.}
\label{fig:3link}
\end{figure*}

In this paper, we consider two fundamental cost functions for systems with inertially-dominated dynamics, corresponding respectively to the kinetic energy the system must have to move at a given speed and to the power losses incurred by the actuators in generating the system's internal forces (which do not always do net work on the system). We then show that when we normalize gaits by fixing the energy or power supplied to the system to a given value, each of these cost functions can be transformed into a geometric relationship between the kinematics of the gait cycle in the shape space and the time taken to execute the gait:
\begin{enumerate}
\item When the cost is taken as the kinetic energy required for the motion, the cycle period $T$ is proportional to the pathlength of the cycle (measured according to a metric derived from the system's inertia), and so scales linearly with the amplitude of the gait cycle.

\item When the cost is taken as the non-regenerable power consumption of the actuators, a good model for this power consumption is the square of the actuator forces, which geometrically corresponds to a squared norm of the system's \emph{covariant acceleration} (the acceleration relative to its unforced trajectories), weighted by the locations where the actuators are attached to the system and integrated over the gait. Under this measure of cost, the time period of a gait is proportional to the \emph{square root} of the amplitude of the gait (because an increase in scale reduces path curvature, allowing the system to move faster at a given acceleration, and thus partially offsetting the effect that increasing inertial pathlength has on the time needed to execute the gait). Additionally, this measure of cost encourages the system to move more slowly in curved sections of its trajectory than in straighter sections (rather than maintaining constant kinetic energy as in gaits that minimize work).
\end{enumerate}
We then use these two cost functions to extend our geometric gait optimization framework (previously developed for swimmers in viscosity-dominated flows~\cite{Ramasamy:2016aa,Hatton:2017aa,Ramasamy:2019aa}) so that it provides optimal gaits for inertia-dominated systems.

Gaits optimizing speed with respect to the kinetic-energy cost function take the same ``soap bubble" form as was seen for the viscous-optimized gaits (maximizing the ratio of signed constraint curvature enclosed by the gait to the pathlength of the cycle used to make this enclosure). As illustrated in Fig.~\ref{fig:InertialIntroFig}, gaits optimizing speed with respect the covariant acceleration cost function tend to be both larger-amplitude than the energy-optimal motions (because the loss terms scale with the square root of amplitude rather than linearly) and rounder than the energy-optimal motions (because the need to slow down for sharp turns adds a ``bending stiffness" to the gait geometry). Together, the energy-optimal and acceleration-optimal gaits bracket the optimal gaits under any cost metric that is weighted sum of the work and loss in the actuators.

As a demonstration of the insights provided by our approach, we apply it to a range of systems, including the isolated three-link system introduced as a representative three-link model in~\cite{walsh95}, the perfect-fluid swimmer from~\cite{Kanso:2005,Melli:2006}, two-mode continuous-curvature swimmers (which we considered in viscous flows in~\cite{Ramasamy:2019aa}), and generalizations of these systems to three and four shape variables (which we also considered for viscous flows in~\cite{Ramasamy:2019aa}).

Finally, as an extension of our core analysis, we consider optimal-gait geometry for systems---such as a three-link system with skates or passive wheels---whose locomotion is dictated by direct nonholonomic constraints, but whose dynamics within these constraints are inertially dominated.

\section{Background}
\label{sec:background}

The geometric framework we use in this paper has its roots in works including~\cite{Shapere:1989a,Murray:1993,walsh95,Kelly:1995,ostrowski_98}, with further development in~\cite{Melli:2006,Avron:2008}. Our treatment below is condensed from a series of papers we have written for the robotics community~\cite{Hatton:2011IJRR,Hatton:2013TRO:Swimming,Hatton:2017TRO:Cartography,Ramasamy:2019aa}, and at a deeper mathematical level, in~\cite{Hatton:2015EPJ}.%

For the purposes of this paper, our focus is on the geometric structure of the system dynamics. Accordingly, we work with the components of these dynamics at a relatively high level of abstraction in the equations, and present their instantiation for specific systems graphically rather than as algebraic expressions (which would run to several pages of trigonometric terms if expanded, even for the three-link swimmer). For worked examples of the kinematics of the $n$-link and continuous systems, see the appendices of~\cite{Ramasamy:2019aa}.

\subsection{Geometric Locomotion Model}
\label{sec:geo}

When analyzing a locomoting system, it is convenient to separate its configuration space $\bundlespace$ (i.e.,\ the space of its generalized coordinates $\bundle$) into a position space $\fiberspace$ and a shape space $\basespace$, such that the position $\fiber\in\fiberspace$ locates the system in the world, and the shape $\base\in\basespace$ gives the relative arrangement of the particles that compose it.\footnote{In the parlance of geometric mechanics, this assigns $\bundlespace$ the structure of a (trivial, principal) \emph{fiber bundle}, with $\fiberspace$ the \emph{fiber space} and $\basespace$ the \emph{base space}.}  For example, the positions of both the articulated and continuous-curvature swimmers in Fig.~\ref{fig:3link} are the locations and orientations of their centroids and mean orientation lines, $\fiber = (x,y,\theta)\in SE(2)$. The shapes of the articulated swimmers are parameterized by their joint angles, $\base = (\alpha_{1},\alpha_{2})$ for the three link swimmer and $\base = (\alpha_{1},\alpha_{2},\alpha_{3})$  for the four-link swimmer. The shapes of the continuous curvature swimmers can be described by a set of modal amplitudes multiplied by their curvature modes. In the serpenoid and piecewise-continuous systems, the shape parameters $\alpha$ are weighting functions on curvature modes $\kappa$ defined along the body,\footnote{Body curvature is the first of four kinds of curvature that appear in this paper.} as discussed in~\cite{Hatton:2017TRO:Cartography}.\footnote{In keeping with previous convention, we use $\base$ when discussing the shape in abstract, and $\alpha$ for the specific parameterization of the shape space that is joint angles or modal amplitudes.}

The dynamics of an isolated-inertial or perfect-fluid locomoting system are dictated by its body-frame inertia matrix $M$, which relates the system's kinetic energy $\kineticenergy$ and momentum $p$ to its velocity  as
\beq \label{eq:kineticenergy}
\kineticenergy = \frac{1}{2} \begin{bmatrix} \transpose{\bodyvel} &  \transpose{\basedot} \end{bmatrix} M(\base)  \begin{bmatrix} \bodyvel \\ \basedot \end{bmatrix}
\eeq
and
\beq \label{eq:momentum}
\begin{bmatrix} p_{\fiber} \\ p_{\base} \end{bmatrix} = M(\base) \begin{bmatrix} \bodyvel \\ \basedot \end{bmatrix}.
\eeq
in which $\bodyvel=\fiber^{-1}\fiberdot$ is the body velocity of the system (i.e., $\fiberdot$ expressed in the system's local coordinates), $\base$ is its current shape, $\basedot$ is the rate at which it is changing shape, $p_{\fiber}$ is its momentum through the world in the body-frame directions, and $p_{\base}$ is its momentum in the shape directions.

For an isolated $n$-link system, this inertia matrix can be constructed by pulling back the links' individual inertia matrices $\mu_{i}$ through the Jacobians of the links,
\beq
M (\base)= \sum_{i} \transpose{\jac_{i}(\base)} \mu_{i} \jac_{i}(\base),
\eeq
where $\jac_{i}$ is the Jacobian that maps the body and shape velocities of the system to the body velocity of the $i$th link,
\beq
\fibercirc_{i} = \jac_{i}\begin{bmatrix} \fibercirc \\ \basedot \end{bmatrix}.
\eeq
(Detailed calculations for this Jacobian are provided in~\cite{Ramasamy:2019aa}.)

For a system immersed in a perfect fluid, the inertia matrix can be constructed on a similar principle, augmenting the link inertias with their fluid-added-mass matrices before pulling them back through the link Jacobians~\cite{Kanso:2005}. For continuously-deformable systems, the summation over links is replaced with an integral along the body,
\beq
M(\base) = \int_{\text{body}} \transpose{\jac}(\base,\ell)\ \mu(\ell) \ \jac(\base,\ell) \ d\ell,
\eeq
where $\mu(\ell)$ now refers to the infinitesimal mass and moment of inertia of the portion of the body at $\ell$ (and can be augmented with fluid added mass as in the discrete-link case).

As described in~\cite{Kanso:2005,Hatton:2013TRO:Swimming}, the inertia matrix $M$ can be decomposed into blocks that respectively encode the system's inertia with respect to pure position motion with fixed shape, pure shape motion at a fixed position, and the coupling terms that appear when the system is moving in both shape and position,
\beq
M = \begin{bmatrix} M_{\fiber\fiber} & M_{\fiber\base} \\ M_{\base\fiber} & M_{\base \base}\end{bmatrix}.
\eeq
If the system starts at rest and forces are applied only through the joints (i.e., we do not ``externally push" the base link), conservation of momentum means that its position-space momentum $p_{\fiber}$ remains zero for all time. Under these conditions, the upper blocks of $M$ encode a constraint on the combinations of $\fibercirc$ and $\basedot$ that the system can achieve,
\beq \label{eq:momentumconstraint}
p_{\fiber} = \begin{bmatrix} M_{\fiber\fiber} & M_{\fiber\base}  \end{bmatrix} \begin{bmatrix} \bodyvel \\ \basedot \end{bmatrix} = \mathbf{0}.
\eeq

\begin{figure}[tbp]
\begin{center}
\includegraphics[width=.45\textwidth]{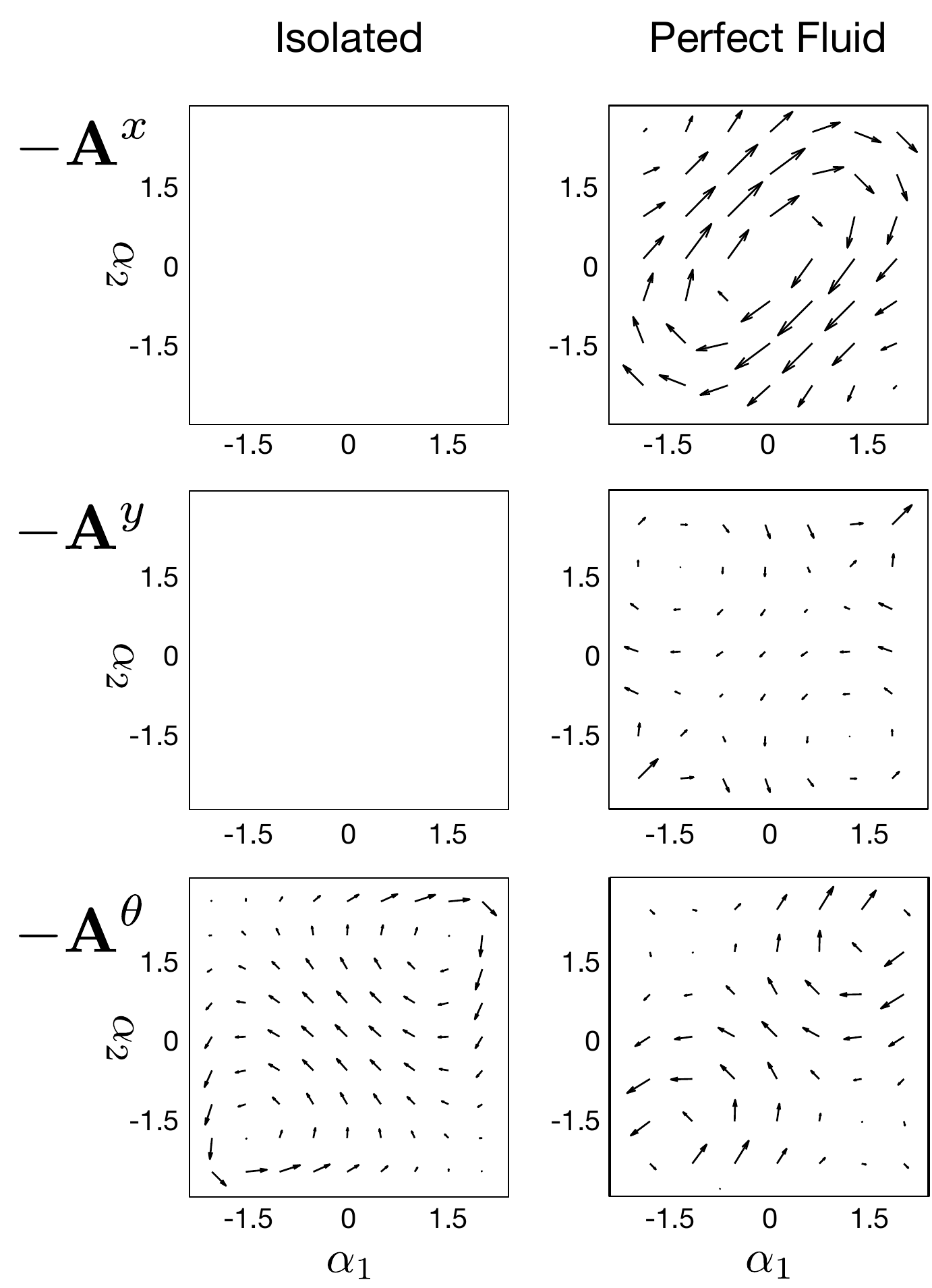}
\caption{The local connection $\mixedconn$ for the isolated and perfect-fluid three-link systems. The $x$ and $y$ fields for the isolated system are zero, because the center of mass (where the body frame for this system is located) cannot move under conservation of linear momentum. Because the added mass on the links is configuration-dependent, conservation of linear momentum does not force the perfect-fluid swimmer to remain stationary, and its $x$ and $y$ fields are non-zero. Both systems have non-zero $\theta$ fields, because their rotational inertias are configuration dependent, so conservation of angular momentum does not prevent the system from changing orientation.}
\label{fig:vecfields}
\end{center}
\end{figure}

If we take the shape velocity $\basedot$ as the ``known" values in this constraint equation, we can reformulate the constraint in~\eqref{eq:momentumconstraint} as a linear map from the shape velocity to the body velocity that puts the system at zero body-frame momentum,
\beq \label{eq:kinrecon}
\fibercirc = -\overbrace{\inv{M}_{\fiber\fiber}M_{\fiber\base}}^{\mixedconn(\base)} \dot{\base}
\eeq
in which the matrix $\mixedconn$ is referred to as the \emph{local connection} $\mixedconn$ for the system, and can be thought of as the mobile-system equivalent to the Jacobian of a robotic manipulator. Each row of $-\mixedconn$ encodes the derivatives of one component of position (in a body frame direction) with respect to the shape components, and can be visualized as an arrow field over the shape space, as illustrated in Fig.~\ref{fig:vecfields} for the isolated and perfect-fluid three-link systems.

\subsection{Gaits} \label{sec:gaits}

Because the shape space of locomoting systems is typically bounded (e.g., by joint limits or other restrictions on bending the body), such systems often move via gaits: cyclic changes in shape that remain within the bounded region of the shape space and produce characteristic net displacements.  Gaits can be described in terms of the \emph{path} the gait traces through the shape space, the \emph{period} required to execute one cycle, and the \emph{pacing} (relative timing) within the cycle.

Several efforts  in the geometric mechanics community~\cite{Murray:1993,walsh95,ostrowski98a,Melli:2006,Morgansen:2007,Shammas:2007,Avron:2008} (including our own ~\cite{Hatton:2013TRO:Swimming,Hatton:2017TRO:Cartography,Ramasamy:2019aa}), have used the \emph{curvature} of the system constraints (a measure of how ``non-canceling" the system dynamics are over periodic shape changes) to understand which gaits produce useful displacements.

The core principle in these works is that because the net displacement $\gaitdisp$ over a gait cycle~$\gait$ is the line integral of~\eqref{eq:kinrecon} along $\gait$, the displacement induced by a gait depends only on the gait's path in the shape space (and not on its period or pacing). Further, the induced displacement can be approximated\footnote{This approximation (a generalized form of Stokes' theorem) is a truncation of the Baker-Campbell-Hausdorf series for path-ordered exponentiation on a noncommutative group, and closely related to the Magnus expansion~\cite{Radford:1998,Magnus:1954vl}. The accuracy of this approximation depends on the body frame chosen for the system, and is most accurate for body frames at an ``average" of the positions and orientations of the body segments; we discuss details of this body frame selection in~\cite{Hatton:2013TRO:Swimming,Hatton:2015EPJ}. In presenting this approximation, we also elide some details of exponential coordinates on Lie groups, which are also discussed in~\cite{Hatton:2015EPJ}.%
} by a surface integral of the constraint curvature\footnote{Constraint curvature is the second kind of curvature to appear in this paper.} $D(-\mixedconn)$ of the local connection (its total Lie bracket\footnote{As discussed in~\cite{Hatton:2015EPJ}, the system motion can be considered as a constrained motion over the full configuration space $\configspace$. The \emph{distribution} of locally-achievable motions can be identified with the vector fields $(\dot{\base}_{i},-\fiber\mixedconn\dot{\base}_{i})$, and the sum of pairwise Lie brackets of these fields over $\configspace$ evaluates to $(0,\extd(-\mixedconn) + \sum[-\mixedconn_{i}, -\mixedconn_{j}])$, in which the latter term is a local Lie bracket on $\fiberspace$ rather than on all of $\configspace$.})   over a surface $\gait_{a}$ bounded by the cycle:
\begin{align}
\gaitdisp &= \ointctrclockwise_{\gait} -\fiber\mixedconn(\base) \label{eq:gaitpathintegral}\\ &\approx \iint_{\gait_{a}} \underbrace{-\extd\mixedconn + \textstyle{\sum}\big{[}\mixedconn_{i},\mixedconn_{j>i}\big{]}}_{\text{$D(-\mixedconn)$ (total Lie bracket)}}, \label{eq:lie}
\end{align}
where $\extd\mixedconn$, the exterior derivative of the local connection (its generalized row-wise curl), measures how changes in $\mixedconn$ across the shape space prevent the net induced motions from canceling out over a cycle, and the local Lie bracket $\sum\big{[}\mixedconn_{i},\mixedconn_{j>i}\big{]}$ measures how sequences of translations and rotations in the induced motions couple into ``parallel parking'' effects that contribute to the net displacement.

For systems with two shape variables, the exterior derivative term evaluates as
\begin{equation}
\extd\mixedconn = \left(\frac{\partial\mixedconn_{2}}{\partial\base_{1}} - \frac{\partial\mixedconn_{1}}{\partial\base_{2}}\right) d\base_1\wedge d\base_2,
\label{eq:extderiv}
\end{equation}
where $\mixedconn_{i}$ is the $i$th column of the local connection (corresponding to the $i$th shape variable), and the local Lie bracket term for planar translation and rotation evaluates as
\beq
\big{[}\mixedconn_{1},\mixedconn_{2}\big{]} = \begin{bmatrix} \mixedconn^{y}_{1}\mixedconn^{\theta}_{2}- \mixedconn^{y}_{2}\mixedconn^{\theta}_{1} \\ \mixedconn^{x}_{2}\mixedconn^{\theta}_{1}- \mixedconn^{x}_{1}\mixedconn^{\theta}_{2} \\ 0 \end{bmatrix}d\base_1\wedge d\base_2.
\label{eq:liebracket}
\eeq
In both cases, the wedge product $d\alpha_1\wedge d\alpha_2$ indicates the oriented differential area basis in the shape space.

\begin{figure}[tbp]
\begin{center}
\includegraphics[width=.45\textwidth]{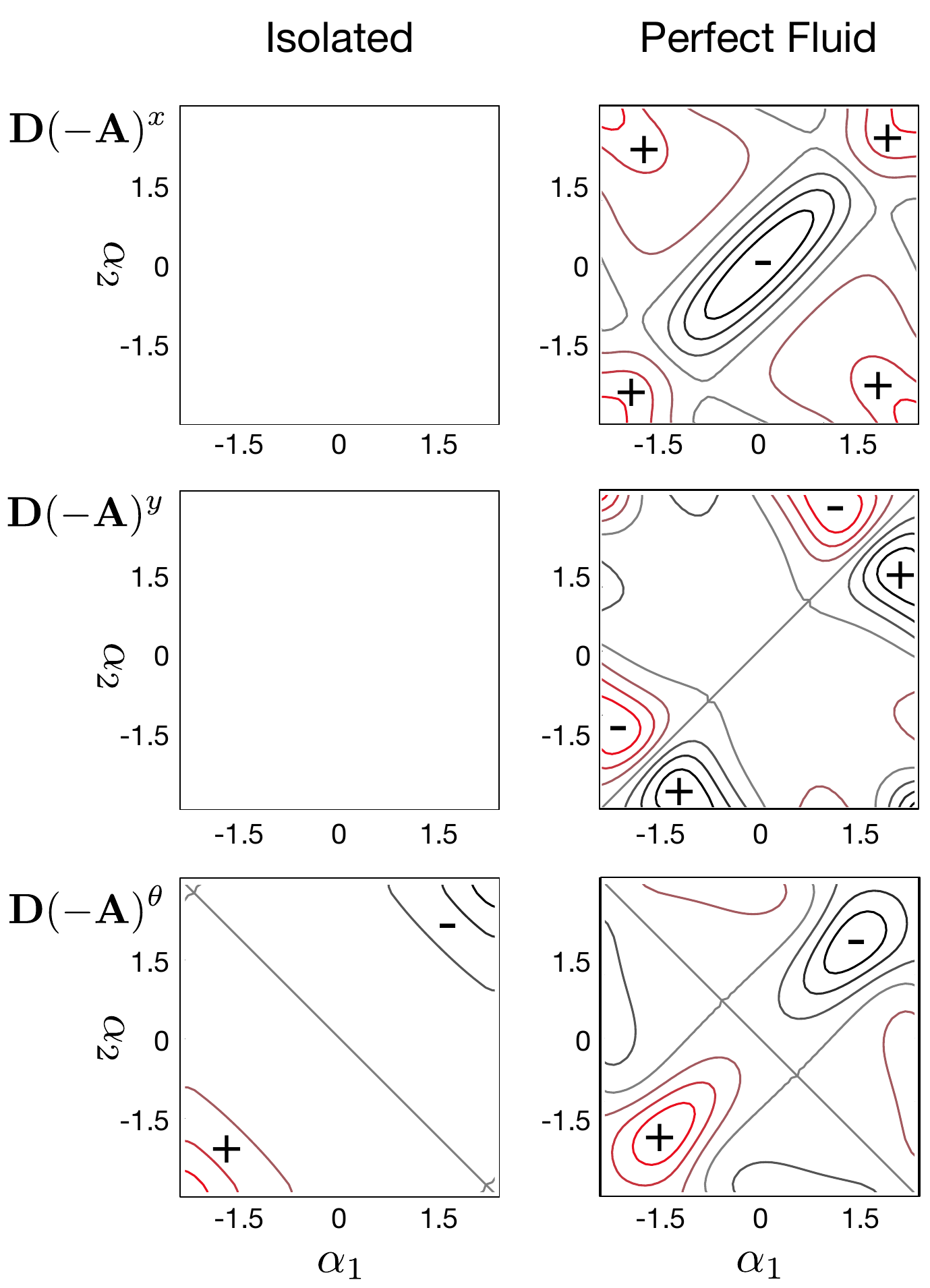}
\caption{The constraint curvature $\covextd{(-\mixedconn)}$ for the isolated and perfect-fluid three-link systems. The $\theta$ components of the constraint curvature are the exterior derivatives (``curls") of the $-\mixedconn^{\theta}$ fields in Fig.~\ref{fig:vecfields}, which captures the ability of the system to make a cycle in its shape space that induces more counterclockwise body rotation than clockwise body rotation. The $x$ and $y$ components of the constraint curvature are the exterior derivatives of their respective $-\mixedconn^{x}$ and $-\mixedconn^{y}$ fields (capturing their ability to execute shape cycles that have more ``forward" motion than ``backward" motion in these body directions, augmented by the local Lie bracket term from \eqref{eq:liebracket}, which captures the ``parallel parking" effects of mixed translation and rotation).%
}
\label{fig:CCFs}
\end{center}
\end{figure}

Plotting these curvature terms as scalar functions over the shape space (as in Fig.~\ref{fig:CCFs}) reveals the effect of gaits' geometry on the motions they induce: Gaits that produce large net displacements in a given $(x,y,\theta)$ direction are located in strongly sign-definite regions of the corresponding $D(-\mixedconn)$ constraint curvature functions (CCFs). For example, $\theta$ rotations of both the isolated and perfect-fluid three-link systems are produced by cycles in the corners of the shape space, whereas cycles centered in the shape space produce net $x$ translations of the perfect-fluid swimmers.

\section{Inertial Dynamics}
\label{sec:inertialdynamics}

The linear map from shape velocity to body velocity in~\eqref{eq:kinrecon} means that the system's kinetic energy, expressed as a function of the body and shape velocity in~\eqref{eq:kineticenergy}, can be expressed entirely as a function of the shape velocity,
\begin{align} \label{eq:kineticenergyreduced}
\kineticenergy &= \frac{1}{2} \transpose{\basedot} \overbrace{\begin{bmatrix} -\transpose{\mixedconn(\base)} &  \matrixid \end{bmatrix} \underbrace{\begin{bmatrix} M_{\fiber\fiber} & M_{\fiber\base} \\ M_{\base\fiber} & M_{\base \base}\end{bmatrix}}_{M(\base)}  \begin{bmatrix} -\mixedconn(\base) \\ \matrixid \end{bmatrix}}^{\displaystyle{M_{\base}(\base)}} \basedot
\\
&= \frac{1}{2} \transpose{\basedot} M_{\base}(\base) \basedot. \label{eq:kineticenergyreduced2}
\end{align}
Note that $M_{\base}$, the \emph{reduced inertia matrix} for the system, is distinct from the $M_{\base\base}$ block of $M$ (which encodes the shape-space inertia for the case where the body frame is fixed, whereas $M_{\base}$ takes the body frame as moving according to the relationship encoded in $-\mixedconn$).

Because $M_{\base}$ encodes the system's full inertial information as a structure on the shape space, it allows us to evaluate the dynamics of the system entirely on the shape space, in what is called a \emph{sub-Riemannian} approach~\cite{Montgomery:2002vn,Ramasamy:2019aa}. We briefly review here some key ideas in evaluating the motion of systems with second-order dynamics, with a focus on the geometry underlying the dynamics of these systems.

\begin{figure*}[tbp]
\begin{center}
\includegraphics[width=\textwidth]{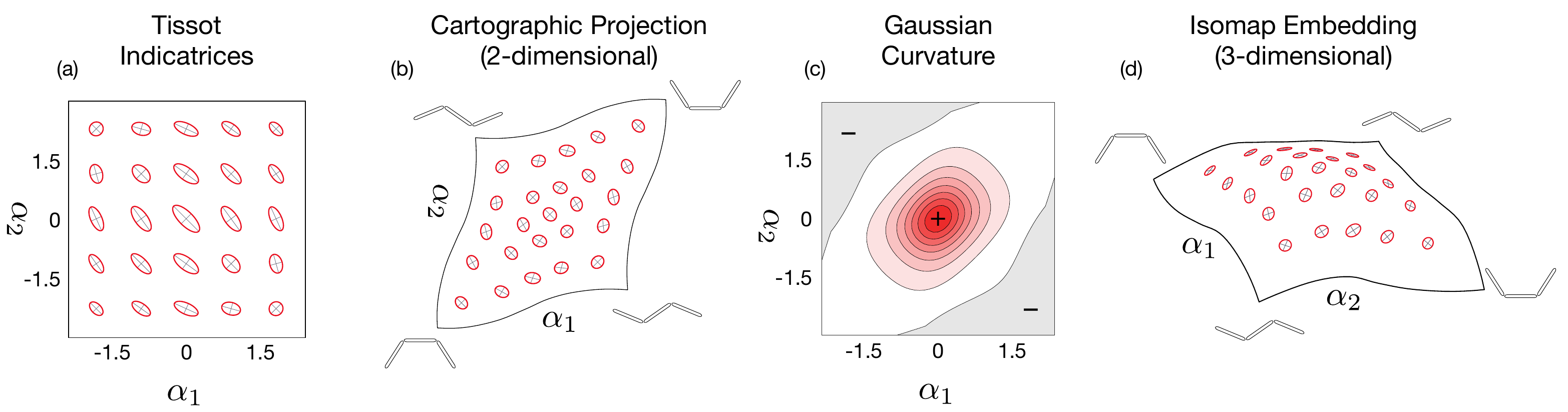}
\caption{The inertia matrices for our systems act as \emph{Riemannian metrics} on their shape spaces. (a) In coordinates, these metrics can be visualized via their Tissot indicatrices (the sets of velocity vectors at each point that produce unit kinetic energy under the metric). (b) To (approximately) recover the geometry of the corresponding inertial manifold, we can use algorithms such as cartographic projection~\cite{Hatton:2017TRO:Cartography} to stretch the shape space so that the indicatrices become as close to uniform circles as is possible while maintaining the continuity of the embedding. The inertial metric illustrated here is specifically that of the three-link isolated system, but it is qualitatively the same as those of the other systems we consider here: The true inertial geometry is stretched relative to the system coordinates along the even axis of the shape space. (c) The structure of the metric means that the inertial manifold has positive Gaussian curvature (i.e., it is domed/cupped rather than saddle-shaped) at the center, and becomes flat (zero curvature) to slightly saddle-shaped (negative curvature) towards the edges. (d) This domed structure can be partially visualized by using an algorithm such as Isomap~\cite{tenenbaum2000} to approximate an isometric embedding of the manifold in three-dimensional space.}
\label{fig:Metrics}
\end{center}
\end{figure*}

\subsection{Shape-space Dynamics}
If we take the kinetic energy expression in \eqref{eq:kineticenergyreduced2} as defining a Lagrangian for the system, the Euler-Lagrange equations dictate a relationship between shape-space forces $\tau$ and motion through the shape space,
\begin{align}
\tau &= M_{\base}(\base)\ddot{\base} + C_{\base}(\base,\basedot) \label{eq:EL},
\end{align}
in which the vector $C_{\base}$ (which encodes the centrifugal and Coriolis forces acting on the system) is calculated from the derivatives of the mass matrix with respect to the shape variables as
\begin{equation} \label{eq:ELC}
C_{\base}(\base,\basedot) = \left(\sum_{i=1}^{d} \parderiv{M_{\base}(\base)}{\base_{i}}\basedot^{i}\right) \basedot
-\frac{1}{2} \begin{bmatrix} \basedot^{T} \parderiv{M_{\base}(\base)}{\base_{1}} \basedot \\ \vdots \\ \basedot^{T} \parderiv{M_{\base}(\base)}{\base_{d}} \basedot \end{bmatrix}.
\end{equation}

In some cases (including later sections of this paper), it is useful to factor the right-hand side of~\eqref{eq:EL} as
\beq \label{eq:covaccel}
\tau = M_{\base}(\base) \overbrace{\bigl( \ddot{\base} + \inv{M}_{\base}(\base)C_{\base}(\base,\basedot) \bigr)}^{\displaystyle{a_{\text{cov}}}},
\eeq
in which $a_{\text{cov}}$, the system's \emph{covariant acceleration}, describes the rate at which it is accelerating relative to its natural (unforced) trajectories.\footnote{The $\inv{M}_{\base}C_{\base}$ term evaluates  to the \emph{Christoffel symbols} associated with $M_{\base}$. We prefer this expression over the formula that directly encodes the Christoffel symbols because it more directly illustrates how the system inertia contributes to the final expression of the dynamics.} The magnitude of the covariant acceleration, taken with respect to the inertia matrix $M_{\base}$ as
\beq \label{eq:covaccelnorm}
\norm{a_{\text{cov}}} = \sqrt{a_{\text{cov}}^{T}\, M_{\base}\, a_{\text{cov}}},
\eeq
is equal to the total (mass-weighted) acceleration of the particles making up the system \emph{that is not due to the constraint forces}, and thus to the total force that must be actively applied to the particles to follow the trajectory.

Taking the magnitude of $a_{\text{cov}}$ with respect to the \emph{square} of the inertia matrix (i.e., taking the ``$\metric^{2}$-norm of the acceleration"), returns a value equal to the Euclidean norm of the shape forces,
\beq \label{eq:afnorm}
\norm{a_{\text{cov}}}_{\tau} = \sqrt{a_{\text{cov}}^{T}\, M^{2}_{\base}\, a_{\text{cov}}} = \sqrt{\tau^{T}\tau}.
\eeq
If the shape space parameterization is chosen such that the shape parameters correspond to the directly-actuated degrees of freedom (e.g., using a joint-angle parameterization for an articulated chain whose joints are individually driven by motors), then $\tau^{T}\tau$ is the sum of squared actuator forces, and $\norm{a_{\text{cov}}}^{2}_{\tau}$ then corresponds to the effort the system actually has to exert to achieve the motion.\footnote{The magnitude of forces applied to the particles is different from the magnitude of the actuator forces because the latter value accounts for the leverage that the actuators have on the masses.
Taking the $\tau$-norm of $a_{\text{cov}}$ is \emph{not} equivalent to replacing the metric $M_{\base}$ with a new metric: the covariant acceleration is still calculated with respect to $M_{\base}$ as in~\eqref{eq:covaccel}, and the $M_{\base}^{2}$ term in~\eqref{eq:afnorm} specifically corresponds to the presence of a \emph{cometric} that is not dual to the metric, and which calls out the actuator forces as being a better measure of effort than the net forces acting on the particles.}\footnote{ Note that although the standard measure of covariant acceleration in~\eqref{eq:covaccelnorm} is invariant under changes of coordinates, the $\tau$-norm in $\eqref{eq:afnorm}$ depends on the choice of shape coordinates (and is physically meaningful because the choice of coordinates is tied here to the physical placement of the actuators on the system, and thus the leverage that they have on the system masses).}

As we discuss in \S\ref{sec:inertialcost}, these shape-space dynamics can be combined with the constraint-curvature analysis in \S\ref{sec:gaits}, which predicts the net displacement induced by a gait from a structure that is also defined on the shape space, to perform a full cost-and-displacement analysis using only shape space structures.

\subsection{Inertial Geometry}
\label{sec:inertialgeometry}

The reduced inertia matrix acts as a \emph{Riemannian metric} on the shape space, defining a weighted two-norm for velocity vectors on the shape space such that the inertially-weighted speed of the system is the square root of twice the system's kinetic energy,
\beq \label{eq:shapevelnorm}
\norm{\basedot}_{M_{\base}} = \sqrt{\transpose{\basedot} M_{\base}(\base) \basedot} = \sqrt{2 \kineticenergy}.
\eeq
This norm allows us to visualize the structure of the reduced inertia matrix by constructing a set of \emph{Tissot indicatrices}~\cite{Tissot:1881,Hatton:2017TRO:Cartography}, as illustrated in Fig.~\ref{fig:Metrics}. The indicatrix at each point in the shape space is the set of $\basedot$ vectors with unit norm ($\norm{\basedot}_{M_{\base}} = 1 $) at that point, which form an ellipse in the corresponding tangent space. The short axes of the indicatrices correspond to configurations and directions in which the system is ``heavy" with respect to the joints, such that less joint motion is required to achieve a given kinetic energy. Conversely, the long axes of the indicatrices correspond to configurations and directions in which the system is ``light" with respect to the joints, such that more joint motion is required to produce a given kinetic energy.

For example, the shape-position velocity coupling means that at a given $\dot{\alpha}$ joint velocity, it takes less kinetic energy for a three-link system to move along the ``odd" ($\mathsf{S}$-shaped) axis of the shape space than it takes to move along the ``even" ($\mathsf{C}$-shaped) axis. As illustrated in Fig.~\ref{fig:Metrics}(a), the Tissot indicatrices (sets of unit-norm velocities) are longer $\dot{\alpha}$ vectors in directions along the odd axis than they are along the even axis.

Some of the true (inertial) geometry of the system can be captured by ``stretching" the shape space~\cite{Hatton:2017TRO:Cartography} so that the Tissot indicatrices become closer to being circles (much in the same way that a good map projection can reduce distortion of map features). For the three-link system, this means stretching along the even axis of the shape space, as illustrated in Fig.~\ref{fig:Metrics}(b), revealing that pairs of $\mathsf{C}$-shapes with a given magnitude of $\alpha$ are inertially further apart from each other than are pairs of $\mathsf{S}$-shapes with corresponding joint angles.

In general, the true inertial geometry is \emph{curved}, and cannot be completely recovered simply by stretching the space (much as the curvature of the Earth means that no flat map can ever be completely distortion-free). The \emph{Gaussian curvature}\footnote{Gaussian curvature of the inertial geometry is the third kind of curvature to appear in this paper.} of the inertial geometry can be calculated from $M_{\base}$ and its derivatives across the shape space via the {Brioschi formula}~\cite{abbena2017modern}.
The Gaussian curvature of the three-link system, illustrated in Fig.~\ref{fig:Metrics}(b) is mostly positive (meaning that the system's inertial geometry is cupped or domed), except for some negative regions at the edges of the space under consideration (meaning that the true geometry is saddle-shaped or wrinkled in these regions). Some of this curvature can be directly visualized by using an algorithm such as Isomap~\cite{tenenbaum2000} to approximate an isometric embedding of the inertial manifold into three-dimensional space, as in Fig.~\ref{fig:Metrics}(c); in most cases, however, more than one extra dimension is required to exactly represent the geometry.

In the subsequent sections of this paper, all of our calculations will be in the natural shape coordinates for each system, using $M_{\base}$ as a set of weights on shape velocities and accelerations. The manifold visualizations, however, play a key role in understanding the results of these calculations and the features we expect to see in the results:
\begin{quote}
\textbf{The dynamics of the inertia-weighted system on the shape space are equivalent to those of a unit point mass constrained to move over the inertia-defined surface.}
\end{quote}
Recognizing this equivalence provides a context and vocabulary for discussing the geometry of the systems' optimal trajectories.

\subsection{Geometry of the Shape-space Dynamics}

An intuitive understanding of the geometry of the shape-space dynamics can be achieved by transforming the system's velocity and covariant acceleration into a local basis constructed such that it is orthonormal with respect to the reduced mass matrix $M_{\base}$ and that the first vector in the basis is in the same direction as $\basedot$.
In this ``primed" frame, the system velocity takes the form
\beq
\basedot' = \begin{bmatrix} v \\ \mathbf{0} \end{bmatrix}
\eeq
where $v = \norm{\basedot}_{M_{\base}} = \sqrt{2\kineticenergy}$ is both the inertial speed of the system and the velocity of the corresponding unit point mass traveling over the inertia-defined surface with the same $\basedot$ parameter velocity.

The inertia-weighted pathlength of a trajectory through the shape space, %
\beq
S = \int_{\gait} \sqrt{\transpose{\basedot} M_{\base}(\base)\basedot}\, dt = \int_{\gait} \sqrt{2\kineticenergy (t)}\, dt = \int_{\gait} v(t) \, dt,
\eeq
is equal to the pathlength of the corresponding trajectory embedded into the inertia-defined surface, and is proportional to the time-integral of the square root of the system's kinetic energy. This inertial pathlength integral can be written without reference to time as
\beq
S = \int_{\gait} \sqrt{\transpose{d\base}\, M_{\base}(\base)\,d\base},
\eeq
meaning that the integrated square root of kinetic energy is a property of the trajectory's path, and is specifically independent of both the period and pacing with which it is followed.

The shortest inertial paths between points in the shape space are \emph{geodesic paths} for the system. The geodesic paths are also the generalization of ``straight paths" within the inertial surface, and the system's natural (unforced trajectories), or \emph{geodesics}, follow the geodesic paths at constant inertial speed.

If the the metric-orthonorma,l path-aligned basis is additionally constructed such that the covariant acceleration lies in the plane formed by the first and second basis vectors, the covariant acceleration takes the form
\beq \label{eq:orthalignedaccel}
a'_{\text{cov}} = \begin{bmatrix} \dot{v} \\ \kappa v^{2} \\ \mathbf{0} \end{bmatrix},
\eeq
where $\dot{v}$ is the rate at which the inertial speed is changing and $\kappa$ is the curvature of the trajectory \emph{within the surface}.\footnote{Trajectory curvature is the fourth and final kind of curvature to appear in this paper.} Together, these components describe the extent to which the system is deviating from an unforced trajectory.

The squared norm of the covariant acceleration, taken in bases orthonormal with respect to the metric tensor, is equal the squared sum of the speed- and direction-change components of the acceleration in the orthonormal bases,
\begin{align}
\norm{a_{\text{cov}}}_{M_{\base}}^{2} &= \transpose{a}_{\text{cov}} M_{\base} a_{\text{cov}} = a'_{\text{cov}} \cdot a'_{\text{cov}} \\[1ex]  &= \dot{v}^{2} + (\kappa v^{2})^{2}. \label{eq:orthaccelnorm}
\end{align}
This norm of the acceleration corresponds to the amount of force that must be applied \emph{tangent to the surface} to push the unit point mass along its trajectory, or equivalently to the magnitude of the vector-sum of non-constraint forces that must be applied to the particles in the system.

In metric-orthonormal coordinates, the squared norm of forces appears as a weighted inner product,
\begin{align}
\norm{a_{\text{cov}}}_{\tau}^{2} &= \transpose{\tau} \tau = \transpose{a'}_{\text{cov}} \overbrace{\begin{bmatrix} b_{\dot{v}}  & b_{v\dot{v}} \\ b_{v\dot{v}} & b_{v} \end{bmatrix}}^{M_{\tau}} a'_{\text{cov}} \\[1ex]  &= b_{\dot{v}}\dot{v}^{2} + 2b_{v\dot{v}}(\dot{v}\kappa v^{2})  + b_{v}(\kappa v^{2})^{2}. \label{eq:orthtorquenorm}
\end{align}
in which $M_{\tau}$ can be constructed as $J^{}_{\perp}\transpose{J_{\perp}}$, where $J_{\perp}^{}$ is the Jacobian mapping from coordinate bases to the path-aligned metric-orthonormal bases. These weighting terms capture the property that  the actuator forces are not expressed in bases orthonormal with respect to $M$ (equivalently, are not applied directly at the particles), but instead are exerted at specific points on the mechanism, with varying amounts of ``leverage" on the system particles.

\begin{figure*}[t]
{\centering
\includegraphics[width=.95\textwidth]{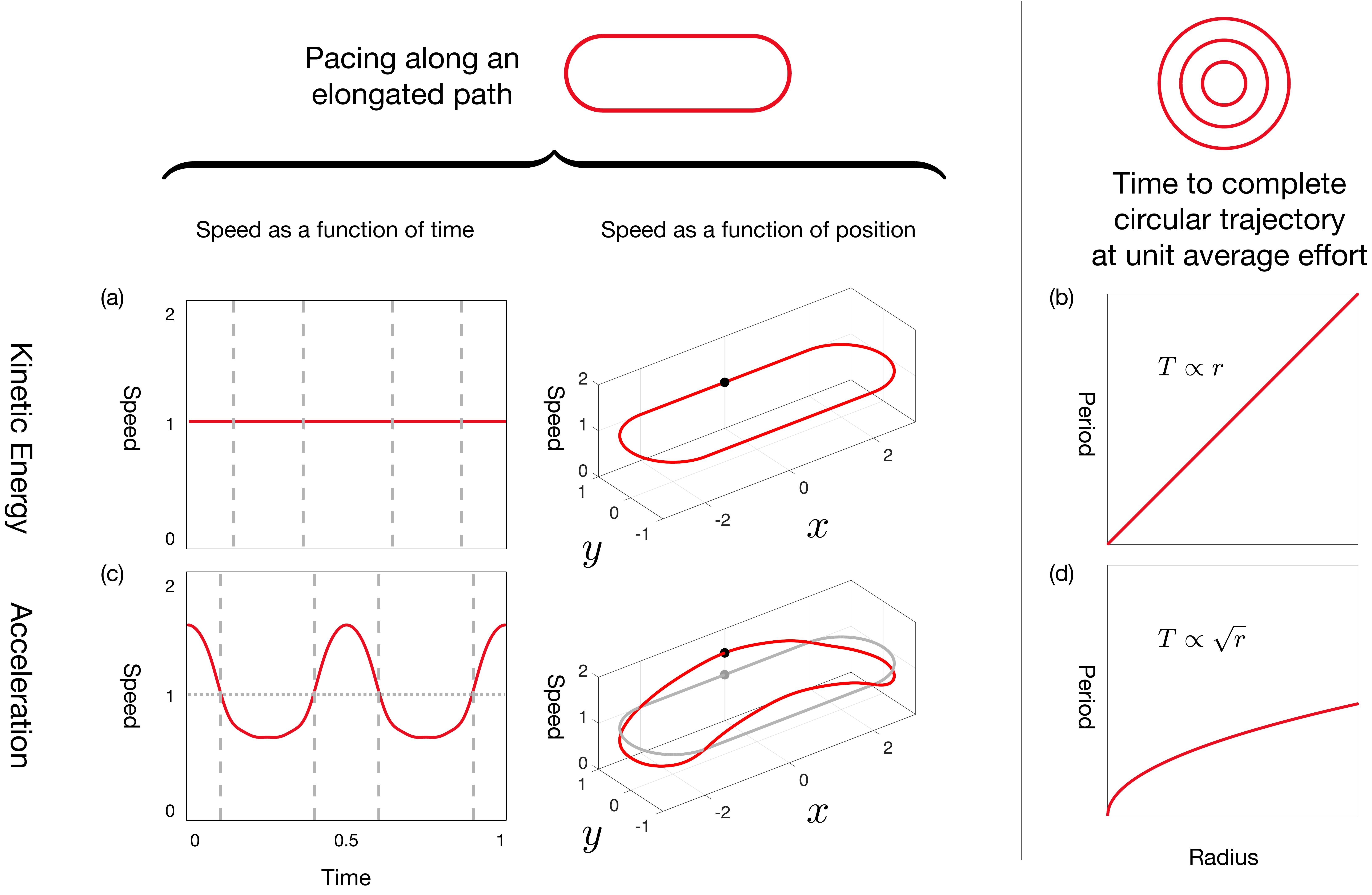}
\caption{Under a kinetic-energy measure of effort, (a) the optimal pacing for a point mass moving along an elongated path is to maintain constant kinetic energy, and (b) the time required for the point mass to move along a circular path at unit kinetic energy is proportional to the radius of the path. If we instead measure the effort required to follow the path as its squared acceleration, (c) the optimal pacing is slower in the curved sections than in the straight sections, and (d) the time required to follow the path at unit acceleration scales with the square root of the path radius.}
\label{fig:racetrack_example}
}
\end{figure*}

\section{Inertial Cost} \label{sec:inertialcost}

The constraint curvature analysis in \S\ref{sec:gaits} provides a clear view of the displacement produced by executing gait cycles, but does not contain any information about the resource or opportunity costs of executing the gaits. This cost information is critical for choosing optimal gaits: a gait that produces a large displacement per cycle but takes a large amount of energy or a very long time to execute is less useful than a gait that produces only a moderate displacement per cycle, but can be executed with a smaller energy expenditure or with higher frequency at the same instantaneous energetic cost, leading to a larger net displacement per energy or time.

Our systems' inertial dynamics, described in \S\ref{sec:inertialdynamics}, suggest two basic means of measuring the effort required for system motion: the kinetic energy required to move with a given shape velocity, and the actuator force required to achieve a given acceleration. %
Integrating these quantities over one cycle of a gait produces an energetic cost that depends on the gait's path through the shape space, its period, and its pacing.

For either effort cost, we can then assign an opportunity cost to each (path, pacing) pair, defined as the period $T$ which, when combined with the path and pacing, produces a gait in which the time-averaged effort cost has unit value. As we discuss in \S\ref{sec:inertialoptimization}, this opportunity cost enables a well-posed gait optimization process, answering the question
\begin{quote}
\textbf{``At a given level of effort, which path and pacing produces the fastest motion through the world?"}
\end{quote}
For the systems considered in this paper, answering this question also answers the question
\begin{quote}
\textbf{``For a given speed through the world, which combination of path, pacing, and period requires the least effort?"}
\end{quote}
but the former question allows for better decoupling between the ``cost" and ``benefit" of a gait (as discussed in \S\ref{sec:inertialoptimization}), and consequently simplifies reasoning about and finding optimal gaits.

Before directly considering the gait optimization problem, we find it useful to examine how the kinetic-energy and actuator-force measures of effort, under the constraint of unit-instantaneous effort, give rise to distinct ``geometries of cost" in the resulting gait period. When combined with the constraint-curvature analysis from \S\ref{sec:gaits}, these geometries of cost provide a complete geometry of optimality for the systems' gaits.

\subsection{Kinetic-energy Cost: Inertial Pathlength}

\label{sec:kineticenergycost}

If we take the system's instantaneous cost of motion as its kinetic energy,
\beq
\inertialcost^{\kineticenergy}_\text{instant} = \kineticenergy,
\eeq
then we can make the following statements about the opportunity costs of gaits:
\begin{enumerate}
\item For any path and mean kinetic energy, the pacing that produces the shortest period $T$ is the one in which the energy is held constant at this mean value. Because kinetic energy defines the metric-normalized speed, this means that the system moves at ``constant speed" with respect to its inertial distribution.
\item Under the constant-inertial speed condition, the period $T$ of a gait is proportional to its metric-weighted inertial pathlength and inversely proportional to the square root of the kinetic energy with which it is executed.
\end{enumerate}

These properties are derived from the following geometric properties of the system dynamics:

\begin{figure*}[t]
{\centering
\includegraphics[width=.95\textwidth]{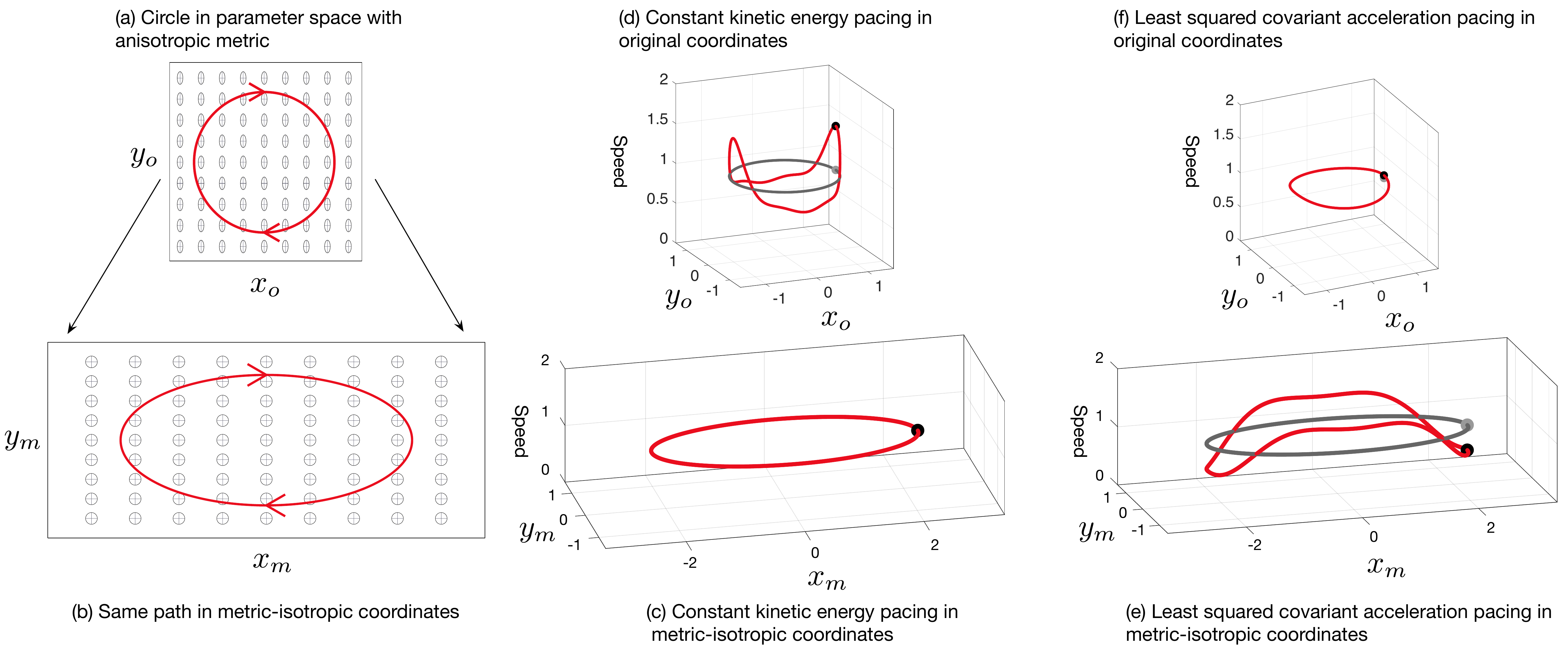}
\caption{If the inertia matrix is not isotropic (e.g., because the system has a shape-dependent distributed mass), then a trajectory that is circular in parameter space, as in (a), may not be truly circular as illustrated in (b). Consequently, the constant-kinetic-energy path in (c) appears to be variable-speed when viewed in coordinates (d) and the squared-covariant-acceleration-optimal pacing for the system (e) may be appear to be constant speed when viewed in coordinates (f), even though the system actually experiences significant tangential acceleration.}
\label{fig:stretch}
}
\end{figure*}

\begin{enumerate}

\item \emph{Optimal pacing:} Because of the quadratic relationship between $\basedot$ and $\kineticenergy$, trading off energy between portions of the gait (moving slower in one portion to move faster in another) at a given mean kinetic energy value always leads to a longer period $T$ than is achieved by maintaining the kinetic energy constant at this mean value: the time gained in the faster sections is necessarily smaller than the time lost in the slower sections. A formal derivation of this principle is presented in Appendix~\ref{app:optpacing}.

\item \emph{Proportionality of period and pathlength:} If the energy in the system is held fixed at a given value, then the inertial pathlength $S$ is equal to the product of the square root of kinetic energy and the gait period,
\beq
S = \int_{0}^{T} \sqrt{2\kineticenergy}\, dt = \sqrt{\kineticenergy}\,T,
\eeq
and the period is correspondingly equal to the inertial pathlength divided by the square root of the kinetic energy at which the gait is executed,
\beq
T = S/\sqrt{2\kineticenergy}.
\eeq
\end{enumerate}

Fig.~\ref{fig:racetrack_example}(a) illustrates the optimal pacing under the kinetic-energy measure of effort for a unit point mass moving around a ``racetrack" path, which is to maintain constant speed through both straight and curved sections of a trajectory. Fig.~\ref{fig:racetrack_example}(b) illustrates the linear proportionality between the radius of a circular path and the time a point mass requires to follow it at unit kinetic energy.

Figs.~\ref{fig:stretch}(a)--(d) illustrate how a non-isotropic inertia matrix affects the pacing in parameters of a constant-kinetic-energy trajectory.
The rate at which gait period scales with respect to geometric scaling of the gait's path (in shape-parameter space) depends both on the coordinate-stretch and on the underlying (coordinate-invariant) curvature of the inertial manifold; as illustrated in Fig.~\ref{fig:covariantacceleration}(a)---(c), gaits for systems with positive Gaussian curvature (``domed" or ``cupped" manifolds) experience slower growth of $T$ with respect to cycle-scaling than systems with flat inertial manifolds, whereas systems with negative Gaussian curvature (``saddle-shaped" manifolds) exhibit increased growth of $T$ with respect to cycle scaling.

Comparing the the system inertia matrix illustrated in Fig.~\ref{fig:Metrics} with the minimal-working-example in Fig.~\ref{fig:stretch}, our systems' inertial geometries are stretched/compressed along the even axis, meaning that during a constant-energy trajectory, the joints move slowly as the system moves from $\rotatebox[origin=c]{180}{\textsf{U}}$ shapes to $\textsf{U}$ shapes, and quickly as it moves from \textsf{S} shapes to \textsf{Z} shapes. The generally-positive Gaussian curvature of the inertial geometry means that the pathlength---and thus period---of gaits can be expected to grow sublinearly with amplitude (which physically corresponds to the system being ``curled more tightly" during higher-amplitude gaits, and thus having a smaller overall moment of inertia).

\subsection{Actuator-force Cost: Covariant Acceleration}\label{sec:accelcost}

The kinetic-energy measure of effort measures the \emph{net} work that the actuators must do on the system to move at a given speed, but does not account for the \emph{individual} work that the actuators perform as they shuffle kinetic energy between different moving pieces of the system. In most cases, this individual work cannot be regenerated or passed between actuators without significant losses, and in some cases actuators may be actively working against each other, with one actuator supplying energy to the system while another dissipates energy.

To more accurately model the effort that the actuators put into the system, we can turn our attention to the actuator forces, which are related to the gait trajectories as in~\eqref{eq:EL}. A complete accounting of the actuator effort to produce those forces requires a detailed model of the actuators, but a good general-purpose model is to take the effort as the squared norm of the actuator forces,
\beq
\inertialcost^{\tau}_\text{instant} = \tau^{T}\tau,
\eeq
capturing the idea that power consumption in an actuator (e.g., the resistive heat losses in an electric motor) grow super-linearly with respect to the force being supplied, and that the costs of producing forces in the individual actuators are decoupled.\footnote{If the cost of producing actuator force is not decoupled, or differs between actuators, this product could be further weighted to reflect such coupling or differing cost.}

As noted in~\eqref{eq:covaccel}, the actuator forces are the product of the system's inertia matrix and its covariant acceleration, such that the squared norm of actuator forces can be expressed in terms of the gait geometry (path and pacing) via the $\metric^{2}$ norm of the covariant acceleration in~\eqref{eq:afnorm}. This proportionality of effort and acceleration results in a relationship between gait path, period, and pacing with the following properties:
\begin{enumerate}
\item For any path and mean squared covariant acceleration, the pacing that produces the smallest period is one in which the system moves slowly in highly-curved sections and faster in straight sections (whereas under kinetic-energy measure of effort, the optimal pacing has constant inertial speed).
\item For any path and mean squared torque, the pacing that produces the smallest period is biased from the acceleration-optimal pacing by a term that depends on the placement of the actuators on the system.
\item The period required to execute a gait with unit average effort is proportional to the fourth root of the effort required to execute the gait in unit time.
\item For geometrically-similar gaits executed at optimal pacing, the period at unit-effort scales with the \emph{square root} of the size of the gait (whereas period under the kinetic-energy measure of effort scales linearly with the size of the gait).
\item For gaits with the same inertial pathlength and mean squared torque, a ``round" gait requires a shorter period than an ``oblong" gait or one with ``sharp corners" (whereas period under the kinetic-energy measure of effort is independent of the aspect ratio or distribution of curvature).
\end{enumerate}

The properties are derived from the following geometric properties of the system dynamics:
\begin{enumerate}[listparindent=1.5em]
\item %
\emph{Acceleration-optimal pacing:} Using the metric-orthornormal representation of the acceleration norm from~\eqref{eq:orthaccelnorm},
the integral of squared covariant acceleration can be written as
\beq \label{eq:localaccelint}
\inertialcost^{a}_{\text{total}} = \int_{0}^{T} \bigl( \dot{v}^{2} + (\kappa v^{2})^{2}\bigr)\, dt.
\eeq
The quadratic nature of this cost function means that as compared to moving with constant $v$, there necessarily exists a pacing that reduces the average acceleration by slowing down the system in sections where $\kappa$ is large, even though this means accepting a non-zero contribution to acceleration from the $\dot{v}$ term. A formal derivation of this principle is presented in Appendix~\ref{app:optpacingaccel}.

\item \emph{Torque-optimal pacing:} In metric-orthonormal coordinates, the squared-torque cost is the integral of the torque-norm of acceleration from~\eqref{eq:orthtorquenorm},
\beq
\label{eq:localtorqueint}
\inertialcost^{a}_{\text{total}} = \int_{0}^{T} \bigl(b_{\dot{v}}\dot{v}^{2} + 2b_{v\dot{v}}(\dot{v}\kappa v^{2})  + b_{v}(\kappa v^{2})^{2} \bigr) \, dt
\eeq
This cost is weighted and biased relative to that in~\eqref{eq:localaccelint}, but has qualitatively similar behavior: its quadratic nature means that relative to a constant-speed profile, shifting acceleration out of the $(\kappa v^{2})^{2}$ and into a term with a $\dot{v}$ term reduces the overall torque cost.

\begin{figure}[t]
{\centering
\includegraphics[width=.45\textwidth]{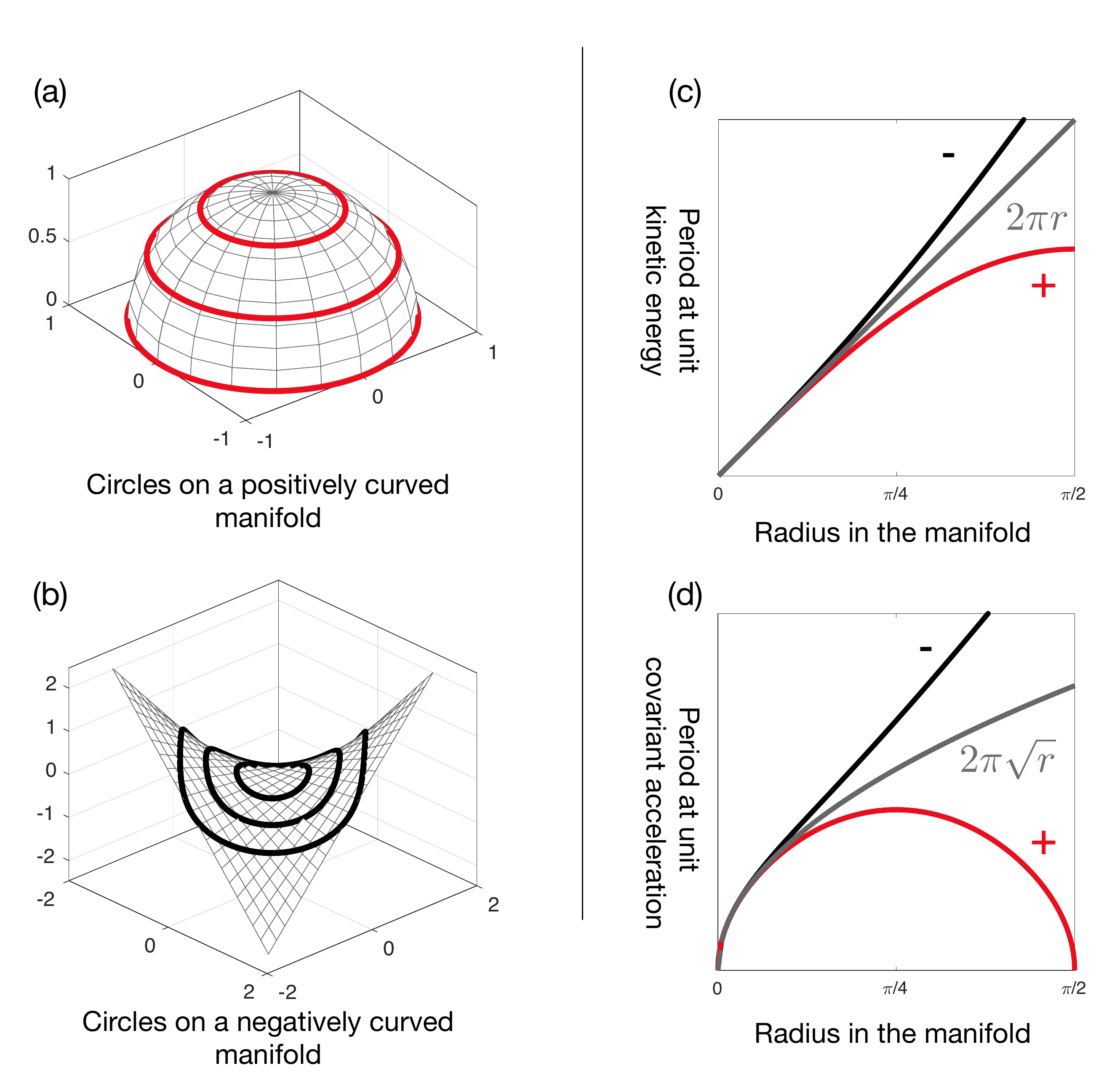}
\caption{The Gaussian curvature of the inertial manifold affects the rate at which gait periods at unit effort grow with respect to the size of the gait. Positive curvature---e.g., the hemispherical geometry in (a)---slows the growth of the period with respect to gait size, and negative curvature---e.g., the saddle geometry in (b)---increases the rate of growth. These rates of growth are illustrated in (c) and (d), with the rate of growth in a ``flat" space provided as reference.
\newline Note that it is possible for a closed path to be a geodesic on the manifold---e.g. following the equatorial great circle on the hemisphere in (a)---in which case the period under the kinetic-energy measure of effort reaches a maximum (because it is proportional to inertial pathlength, and geodesics are curves at extrema of inertial pathlength) and the period under the squared-acceleration measure of effort goes to zero (because geodesics are the trajectories of no acceleration, and the manifold's intrinsic curvature provides all of the ``change of direction" required to close the loop).}
\label{fig:covariantacceleration}
}
\end{figure}

\begin{figure*}[tbp]
\begin{center}
\includegraphics[width=.9\textwidth]{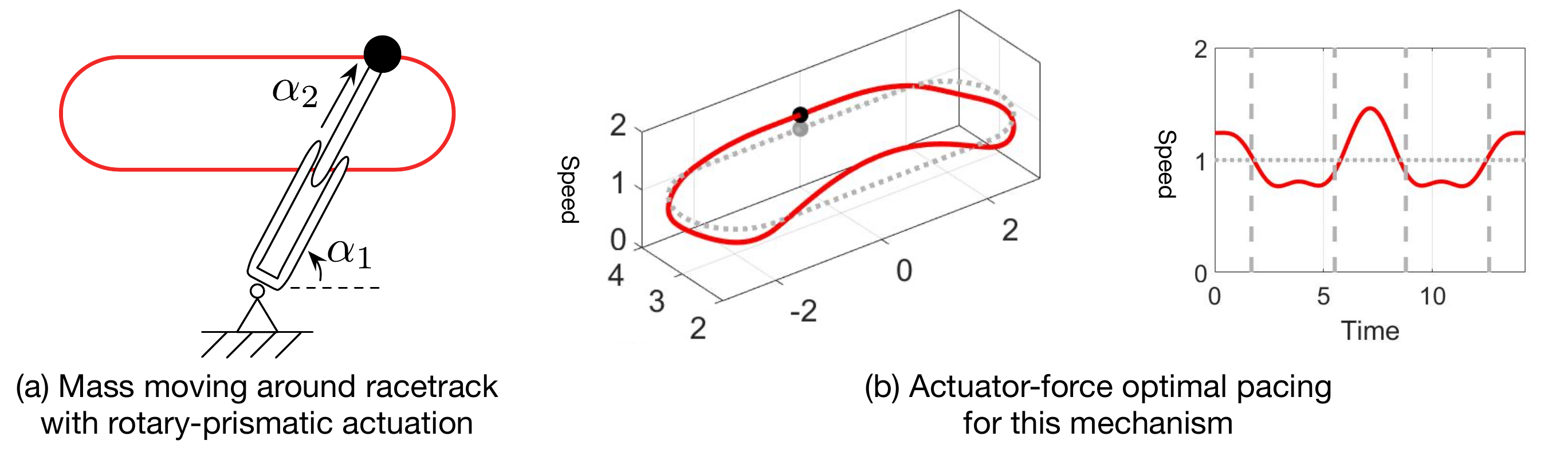}
\caption{Optimizing pacing with respect to squared actuator force is similar to optimizing with respect to covariant acceleration, but introduces a bias term based on the location of the actuators. (a) A point mass being moved around a race-track path by a rotary-prismatic mechanism. (b) The speed of the mass under actuator-force-optimal pacing as a function of both position along the racetrack, (left) and time within the trajectory (right). Note that the mass accelerates more aggressively in the straight section that is closer to the rotary joint than in the section that is further away, because linear acceleration along the track costs less torque when the moment arm is small. Similarly, the force-optimal pacing has a ``speed wobble" during the curves, which correspond to accelerations along the racetrack becoming equivalent to accelerations directly along one of the mechanism joints.}%
\label{fig:rotaryprismaticracetrack}
\end{center}
\end{figure*}

\item \emph{Unit-effort / unit-time proportionality:} Changing the timescale on which the system motions occur by a factor $c$ induces a change the accelerations by a factor $c^{2}$. This factor means that the normed actuator force $\tau_{T}$ during a gait with a given path and pacing and a period $T$ is related to the actuator force during a gait with the same path and pacing but unit period, $\tau_{1}$, by a factor
\beq
\norm{\tau_{T}(t) }= \frac{1}{T^{2}} \norm{\tau_{1}(t/T)}.
\eeq
Combining this relationship with a constraint that restricts our attention to gaits with unit average effort,
\begin{equation}
\label{eq:cost_constraint}
\frac{\int_0^T \norm{\tau_{T}}^{2} dt}{T} = 1,
\end{equation}
tells us that the opportunity cost (period) of a gait with a given path and pacing, executed with unit \emph{average} effort satisfies the relationship
\begin{equation}
\label{eq:cost_constraint_subs}
\frac{\int_0^T \left(\frac{1}{T^{2}} \norm{{\tau_{1}(t/T)}}\right)^{2} dt}{T} = 1.
\end{equation}
Separating out the explicit factors of $T$ in this expression provides an equation
\begin{equation}
T^{5} = \int_0^T  {\norm{\tau_{1}(t/T)}^{2}} dt,
\end{equation}
whose righthand side  can be rewritten to factor out $T$ as
\begin{equation}
\int_0^T  {\norm{\tau_{1}(t/T)}^{2}} dt = T\int_0^1  {\norm{\tau_{1}(t)}^{2}} dt,
\end{equation}
such that the period at unit average effort is revealed as being equal to the fourth root of the effort cost of the gait when it is executed with unit period,
\begin{equation}
\label{eq:time_period_constraint}
T = \left( \int_{0}^{1} \norm{\tau_{1}}^{2} dt \right)^{1/4}.
\end{equation}

\begin{figure*}[tbp]
\begin{center}
\includegraphics[width=.9\textwidth]{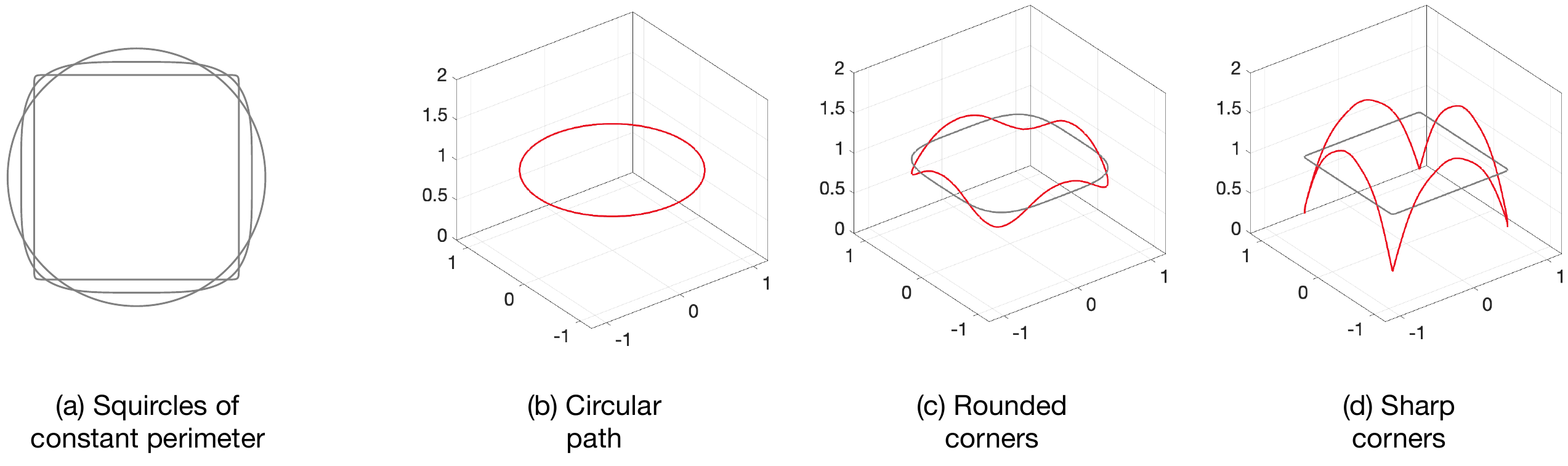}
\caption{For a set of paths with the same perimeter but different distributions of curvature (a), the optimal pacing has more variation in speed the sharper the corners in the trajectory are (b)--(d). These speed changes absorb some of the system's acceleration budget, meaning that it must move slower (and thus take longer) as the curvature distribution along the path becomes more uneven.}%
\label{fig:squircles}
\end{center}
\end{figure*}

\item \emph{Square root proportionality of $T$ and gait amplitude:} Increasing the pathlength of a gait by a factor $\ell$ while maintaining its period, pacing, and curvature distribution scales $v(t)$ by a factor of $\ell$, scales $\dot{v}(t)$ by a factor of $\ell^{2}$, and scales $\kappa(t)$ by $1/\ell$ such that the unit-time torque cost scales by $\ell^{2}$. Inserting this scaling factor into~\eqref{eq:time_period_constraint} then tells us that the unit-effort period for the gait increases by a factor of $(\ell^{2})^{1/4} = \ell^{1/2}$.

Physically, this scaling factor corresponds to the property that as the size of the path increases, it takes proportionally more time to complete a cycle at a given $v$, but the path's curvature \emph{decreases} as the gait grows. The system can thus move with a greater velocity while maintaining the same instantaneous $\kappa v^{2}$ effort in the curves, so that the total time to complete the gait rises sublinearly.

\item \emph{Optimality of roundness:} The quadratic nature of the cost function means that a gait with concentrated curvature (e.g., corners or an oblong shape) incurs greater $(\kappa v^{2})^{2}$ costs at unit period, and therefore requires a greater period at unit effort, than does a system with the same inertial pathlength but more constant (rounder) curvature.

\end{enumerate}

Fig.~\ref{fig:racetrack_example}(c) illustrates how the factors described above influence the optimal pacing of a point mass moving around a ``racetrack" path with unit mean squared acceleration,  and Fig.~\ref{fig:racetrack_example}(d) illustrates the square root proportionality between the radius of a circular path and the time the point mass requires to follow it at unit mean squared acceleration.

As illustrated in Fig.~\ref{fig:stretch}(e), the ``stretch" from an anisotropic inertia matrix can lead to a ``circular" gait in coordinates not being truly circular, such that its optimal pacing under the kinetic-energy metric proceeds around the circle at a non-constant pace. Conversely, the stretch acts to mask the tangential acceleration (and thus effort) required to follow a circular path at mean squared acceleration, as is illustrated in Fig.~\ref{fig:stretch}(f): the stretch means that the acceleration-optimal pace requires slowing down in the ``narrow ends" of the true elliptical gait geometry, but mapping the motion back into the parameter coordinates slows down the motion in the ``straight sections" of the true geometry, such that the acceleration-optimal motion is constant-speed in the parameter space.

As illustrated in Fig~\ref{fig:covariantacceleration}(d), the time required to follow a shape trajectory grows more slowly with loop size on inertial manifolds with positive (domed/cupped)  curvature as compared to on a flat manifold, and grows more quickly on a manifold with negative (saddled) curvature. The reduction or increase in period for an acceleration-normalized trajectory is stronger than that for an energy-normalized trajectory; in the case of a positively-curved manifold, the required period can actually shrink with increased loop size as the loops approach geodesics on the inertial manifold---e.g., the great circle on the hemisphere in Fig~\ref{fig:covariantacceleration}(a).\footnote{This property is described by the Gauss-Bonnet theorem: the total curvature in a closed loop on a surface is equal to $2\pi$ minus the surface curvature enclosed by the loop.}

The difference between optimal pacing under a raw covariant-acceleration cost and optimal pacing under an actuator-force cost is illustrated in Fig.~\ref{fig:rotaryprismaticracetrack}. The point mass at the end of the rotary-prismatic arm has a flat inertial manifold, so its covariant acceleration is its simple acceleration on the track, and this acceleration is equal (up to units) with the force that must be applied to the mass to produce the acceleration. When these forces are projected onto the joint angles, however, horizontal force requires increased $\alpha_{1}$ torque when $\alpha_{2}$ is large. This biasing term means that the optimal pacing involves a more aggressive acceleration during the ``near" straight section, and also introduces a ``speed wobble" to the curved sections based on how the mass's acceleration becomes intermittently aligned with one or other of the mechanism joints.

Fig.~\ref{fig:squircles} illustrates the effect of roundness on optimal gait speed: For paths of equal perimeter on a continuum between circles and squares, the optimal trajectory pacings slow down at the corners (high curvature regions) to avoid incurring large centripetal acceleration costs. These decelerations mean that the system moves at a lower average speed than it moves through the circular path, thus requiring increased time to traverse non-round paths.

For the locomoting systems we are considering, the cancelation between true-geometry acceleration and coordinate-perceived speed from Figs.~\ref{fig:stretch}(e) and (f) means that the optimal pacing of a gait can be inferred directly from the ``roundness" of its path in shape coordinates. The generally positive Gaussian curvature of the inertial surfaces means that the time taken to execute a gait at a given mean acceleration grows slower than the square root of the radius of the gait.

The torque bias appears as a metric-squared acceleration cost, which means that accelerations along the even axis of the shape space become more costly, biasing the system towards constant-speed motion when moving along the even directions of the shape space and towards straighter motions along the odd axis of the shape space (so as not to have curvature in an even direction).

\section{Optimal Gaits} \label{sec:inertialoptimization}

Combining the geometric relationship between a gait's path and the displacement it produces (from \S\ref{sec:gaits}) with the geometric relationship between its path and pacing and its period at unit average effort (from \S\ref{sec:inertialcost}) provides a geometric relationship between its path and pacing and its efficiency $\eta$, measured as its speed at unit effort,
\beq \label{eq:gaitefficiency}
\eta(\text{path},\text{pacing}) = \frac{\fiber_{\gait}(\text{path})}{T(\text{path},\text{pacing})},
\eeq
where, for conciseness of notation, we now use $\fiber_{\phi}$ to indicate a signed norm of the displacement resulting from the gait (e.g., net rotation, or translation in a specific body direction).

The gradient of this efficiency with respect to a set of gait parameters $p$ that define the path and pacing of the cycle is
\beq \label{eq:gradient1}
\nabla_{p} \eta = \frac{1}{T} \nabla_{p} \fiber_{\gait} - \frac{\fiber_{\gait}}{T^2} \nabla_p T,
\eeq
which can be intuitively described as being the gradient of the net displacement with respect to the parameters, minus the gradient of the cost with respect to the parameters, with a normalizing factor to account for the different units of displacement and cost.

\begin{figure*}[tbp]
\begin{center}
\includegraphics[width=.95\textwidth]{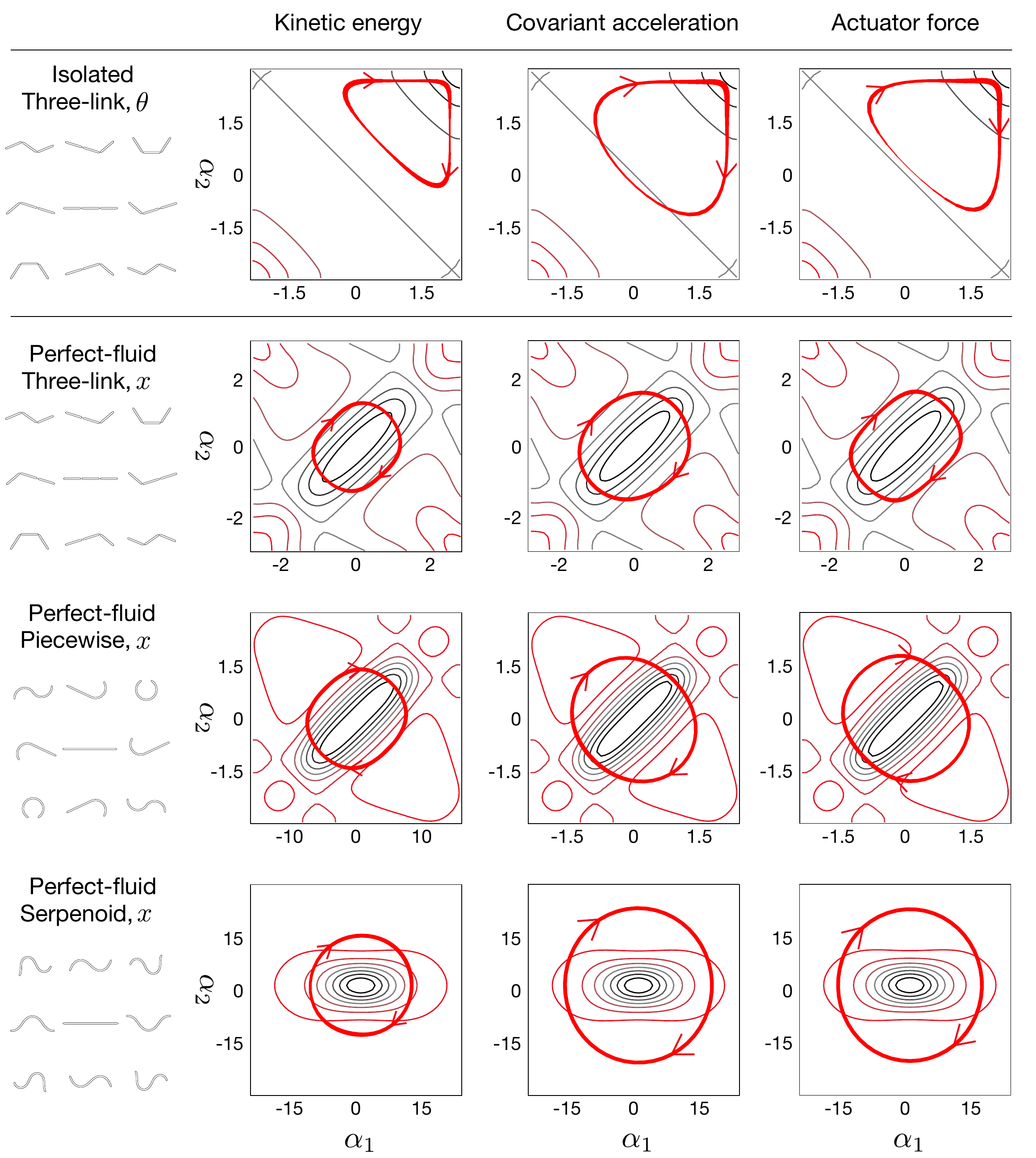}
\caption{Optimal gaits under the kinetic energy, covariant acceleration, and actuator-force metrics for a range of example systems. The optimal pacing is indicated via line thickness, with thick lines indicating ``slow" changes in the shape variables and thin lines indicating ``fast" changes (as if a set of points evenly distributed in time were ``bunched up" or ``stretched out"). The key trends to note are that the gaits optimized for kinetic-energy are universally smaller and less rounded than the gaits optimized for covariant acceleration or actuator-force, and that the actuator-force gaits are shaped and paced so that they have large fast-moving segments in the ``odd" dimension of the shape space (most visible on the isolated three-link and perfect-fluid piecewise gaits). This latter trend corresponds to the squaring of the inertia matrix in the actuator-force calculation, which reinforces the already existing property that motion along the odd axis is ``easier" than motion along the even axis.}
\label{fig:optgaits}
\end{center}
\end{figure*}

\begin{figure*}[t]
{\centering
\includegraphics[width=.95\textwidth]{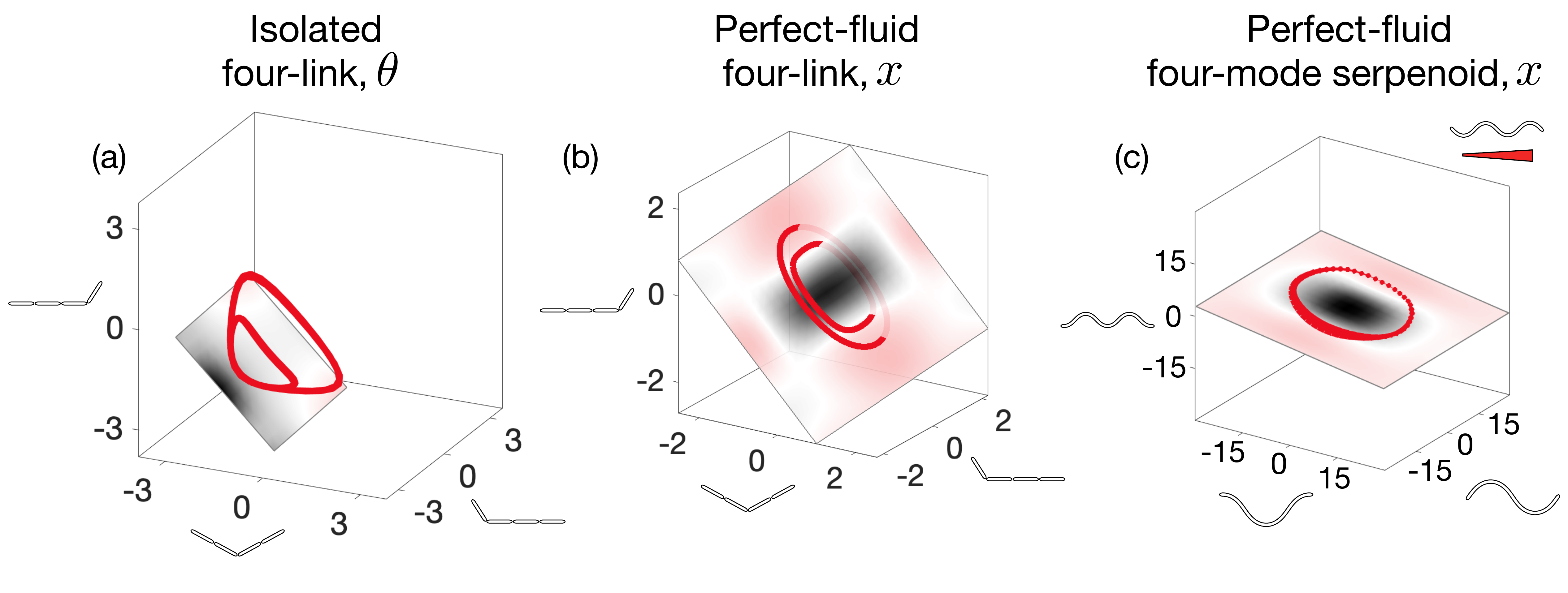}
\caption{Optimal gaits for systems with more than two shape variables. In the first two subfigures, the smaller gait is optimal with respect to kinetic energy and the larger gait is optimal with respect to actuator force; in the third subfigure, the two gaits follow similar enough paths to be visually indistinguishable. The colored planes are positioned such that they pass through the points where $\covextd{(-\mixedconn)}$ is largest, and aligned so as to capture the most flux at those points. The colormaps are the CCFs indicating the flux of the relevant component of $\covextd{(-\mixedconn)}$ through the planes. The optimal gaits for the systems are closely aligned with these planes and enclose rich, sign-definite areas within the planes (as if we had performed a two-dimensional optimization in the set of shape modes defined by the plane). The excursions of the optimal gaits from the planes are to more fully capture flux from $\covextd{(-\mixedconn)}$, which, being a fully three- or four-dimensional differential form changes direction and value across the space and is not completely characterized by the illustrated plane. For the four-mode serpenoid system, the motion of the fourth shape variable is indicated via line thickness.}
\label{fig:highdimshape}
}
\end{figure*}

This gradient serves two purposes in our analysis. First, the gradients of $\fiber_{\gait}$ and $T$ can both be calculated (in closed or semi-closed form) in terms of the constraint curvature $\covextd(-\mixedconn)$ and the inertia matrix $M$.\footnote{See~\cite{Ramasamy:2019aa} for calculation of the gradient of $\fiber_{\gait}$ and the gradient of the inertial-pathlength instantiation of $T$, and Appendix~\ref{app:actuatorforcetimegradient} of this paper for the gradient of the actuator-force instantiation of $T$.} Gradient-ascent optimization in the direction identified in~\eqref{eq:gaitefficiency} then allows for fast optimization of a high-density parameterization of the gait. Second, the gradient identified in~\eqref{eq:gaitefficiency} represents a fundamental truth about the system dynamics, capturing the \emph{fundamental underlying structure} of any other optimization approach applied to these systems.

Under both the kinetic-energy and actuator-force measures of cost, the optimal gaits represent stable equilibria between the $\nabla_{p} \fiber_{\gait}$ and $\nabla_p T$ terms. As we discussed in~\cite{Ramasamy:2019aa}, these equilibria (and thus the optimization process) resemble those defining the shape of soap bubbles---the $\nabla_{p} \fiber_{\gait}$ term acts as an ``inflating pressure" on the size of the gait, pushing it towards the maximum-displacement cycles (which follow zero-contours of the constraint curvature, fully-enclosing sign-definite regions). The pathlength-cost then acts as a ``surface tension" (and for, the force metric, a ``bending strain") term that pushes the gait to give up low-yield regions of the CCFs in favor of shorter (in the sense of both inertial pathlength and period) gaits that can be repeated more often around ``richer" regions of the CCFs.%

\subsection{Optimal Gaits Under the Kinetic-energy Metric}

Under the kinetic-energy metric, the ``surface tension" effect of the boundary cost acts as a true surface tension, penalizing gaits with a long inertial pathlength, but putting no direct penalty on curvature of the path. The way in which the kinetic-energy cost favors moving at constant inertial speed acts like the ``concentration gradient" term in a soap bubble, which enforces an even distribution of material over the surface of the bubble. Optimal gaits under the kinetic-energy metric for the isolated three-link system and a set of perfect-fluid swimmers are illustrated in the first column of Fig.~\ref{fig:optgaits}.

\subsection{Optimal Gaits Under the Actuator-Force Metric}
The optimization dynamics under the actuator-force measure of effort are similar to those under the kinetic-energy cost function, but with the following changes:
\begin{enumerate}
\item The ``surface tension" is now based on the square root of the perimeter (rather than being directly to proportional to it), so that the optimal gaits under the actuator-force metric are larger than those under the energy metric.
\item There is now a ``bending stiffness" on the boundary in addition to the surface tension, such that optimal gaits under the actuator-force metric are rounder than those under the kinetic energy metric.
\item The pacing is now encouraged to slow down in more curved areas, rather than moving around the path at a constant inertial speed. Physically, this pacing change corresponds to constructing the gait curve from a material that contracts axially under applied bending load, producing a higher material density in curved regions.
\end{enumerate}

Optimal gaits for the example systems under both the covariant-acceleration cost metric (which does not account for actuator placement) and the actuator-force metric (which does account for actuator placement) are illustrated in the second and third columns of Fig.~\ref{fig:optgaits}.

\begin{figure*}[t]
{\centering
\includegraphics[width=.85\textwidth]{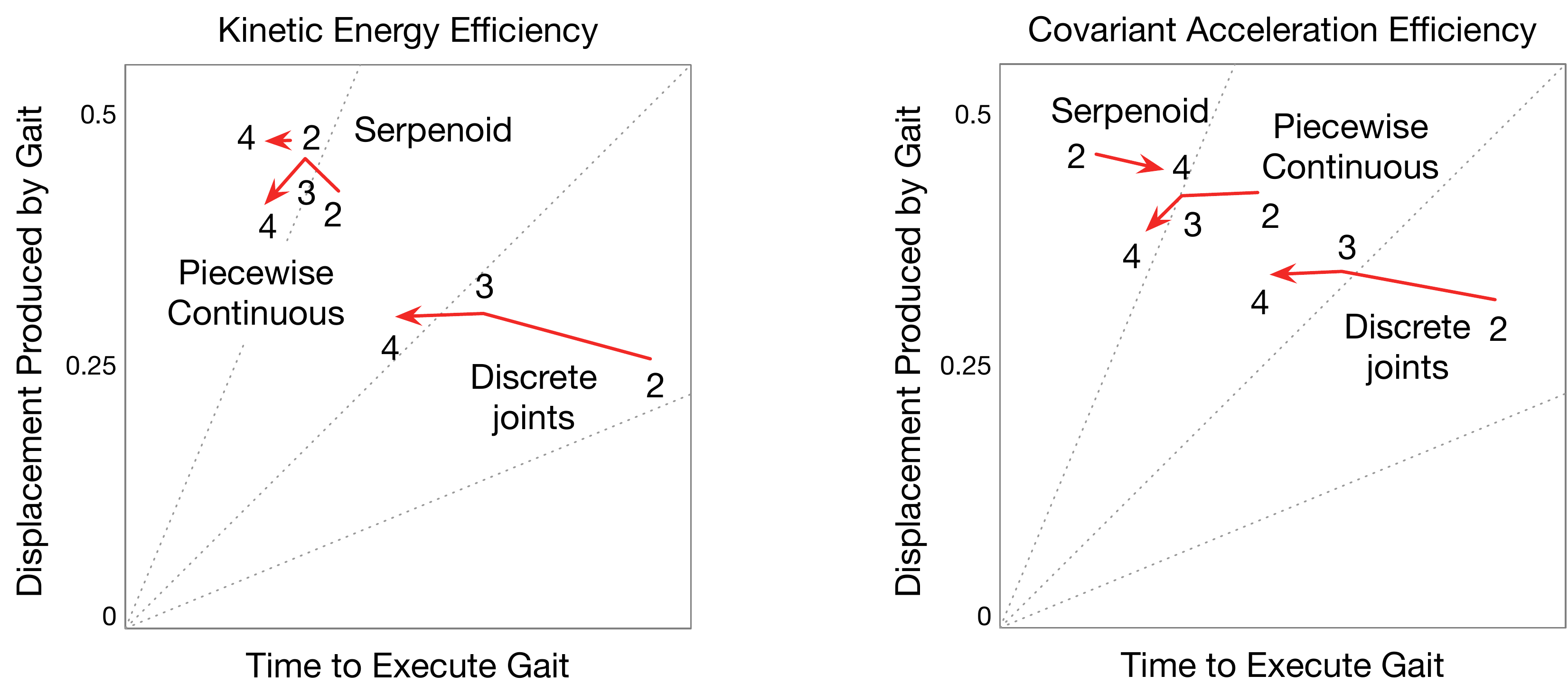}
\caption{Comparison of the most efficient $x$-translation gaits for discrete-joint, piecewise-continuous, and serpenoid swimmers, measured as the displacement they produce (in body lengths) divided by the time required to execute them under the constraints of unit kinetic energy and unit average covariant acceleration. The steeper the line from the origin to the system, the more efficient its optimal gait is. In general, efficiency increases with both mode count and continuity of the modes; the key exception is for the serpenoid swimmer, whose efficiency with respect to covariant acceleration decreases when higher-order modes are added (because the new modes do not significantly increase the capability of the system, but it takes effort to avoid moving along them). Note that the vertical scales on the two plots are comparable, but the horizontal scales are not (because there is no inherent relationship between ``unit kinetic energy" and ``unit average covariant acceleration"). The dotted lines in the plots serve as grid-lines or references; all points on any of these lines are equivalent in terms of displacement-per-time at unit-average-cost, and gaits above/left of a given line are more efficient than gaits that are below/right that line.}
\label{fig:EfficiencyComparison}
}
\end{figure*}

\subsection{Higher-dimensional shape spaces}

As we discuss in more detail in~\cite{Ramasamy:2019aa}, the ``soap bubble"  descriptions of optimal gaits under the kinds of cost functions we discuss here extend directly to systems with more than two shape variables, with the key difference in higher dimensions being how we handle the curvature of the constraints.

In the case of system with three shape variables, the constraint curvature functions can be considered as three-dimensional vector fields, and the displacement of the system over a gait corresponds to the net flux of that field through the loop formed by the gait. The cost of executing a gait remains either its inertial pathlength or the squared norm of the actuator forces, and the optimal gaits are those that ``catch the most flux" relative to the boundary cost.

The structure of a three-dimensional flux field is difficult to visualize in a two-dimensional format, so, as discussed in~\cite{Ramasamy:2019aa}, we find it convenient to represent the $\covextd{(-\mixedconn)}$ fields via projection slices as in Fig.~\ref{fig:highdimshape}: %
For each system, we construct a surface that passes through the point where the magnitude of the relevant $(x,y,\theta)$ component of $\covextd{(-\mixedconn)}$ is largest,  and aligned the surface so that it is normal to that component of $\covextd{(-\mixedconn)}$ at that point. Optimal gaits can be expected to lie close to this plane and encircle sign-definite regions, but bend out of the plane to fully capture the three-dimensional flux

Optimal gaits for four-link (three joint) isolated and perfect-fluid systems are illustrated in Fig.~\ref{fig:highdimshape}(a) and~(b).  For the isolated system, the flux increases towards the corner of the shape space, and the resulting gaits thus force themselves into the corner as much as possible (similarly to the isolated-system gaits in Fig.~\ref{fig:optgaits}). On the perfect-fluid system, the optimal gaits lie very close to the projection plane, but curl slightly outward to maximize the flux they capture from structure not reflected in the plane.

Beyond three dimensions, the CCFs cannot be treated as vectors, but instead must be treated in a fully differential-geometric fashion as ``differential two-forms". At a practical level, this just means that the flux of $\covextd{(-\mixedconn)}$ through the gait loop is directly associated with a surface bounded by the loop, rather than a vector field passing through such a surface,\footnote{See~\cite{Ramasamy:2019aa} for further discussion of this point.} and the geometry of optimal gaits remains the same: they capture the most $\covextd{(-\mixedconn)}$ flux for the least boundary cost. Optimal gaits for a perfect-fluid system with four serpenoid modes are illustrated in Fig.~\ref{fig:highdimshape}(c).

\subsection{System Comparison}

As illustrated in Fig.~\ref{fig:EfficiencyComparison}, efficiency under both the kinetic-energy and covariant-acceleration costs generally increases as the swimmer becomes more continuous (gains extra deformation modes or adopts smoother deformation modes), following the same pattern we observed for the viscous swimmers in~\cite{Ramasamy:2019aa}.

The exception to this rule is the serpenoid swimmer, whose efficiency with respect to kinetic energy barely changes when the second pair of modes is added, and whose efficiency with respect to covariant acceleration actually decreases when the additional modes are added. Referring to the optimal gaits plotted in Fig.~\ref{fig:highdimshape}(c), we can explain this behavior by noting that the optimal gaits lie almost in the plane of the original modes, such that there are no strong kinematic benefits from moving with the additional modes. For the kinetic-energy metric, the presence of the extra modes does not incur any additional costs (because the optimal gait simply does not move along these directions). For the covariant-acceleration metric, however the additional modes do incur additional cost (because the system must actively brace against the natural accelerations in those directions).

We have not plotted relative efficiency under the torque-based cost, because a fair comparison would require construction-specific details about the system (as opposed to the two costs that we do plot, which depend only on the physical geometry of the system and are independent of the actuator placement and internal workings).

\section{Direct Nonholonomic Constraints}

\begin{figure*}[tbp]
\begin{center}
\includegraphics[width=.95\textwidth]{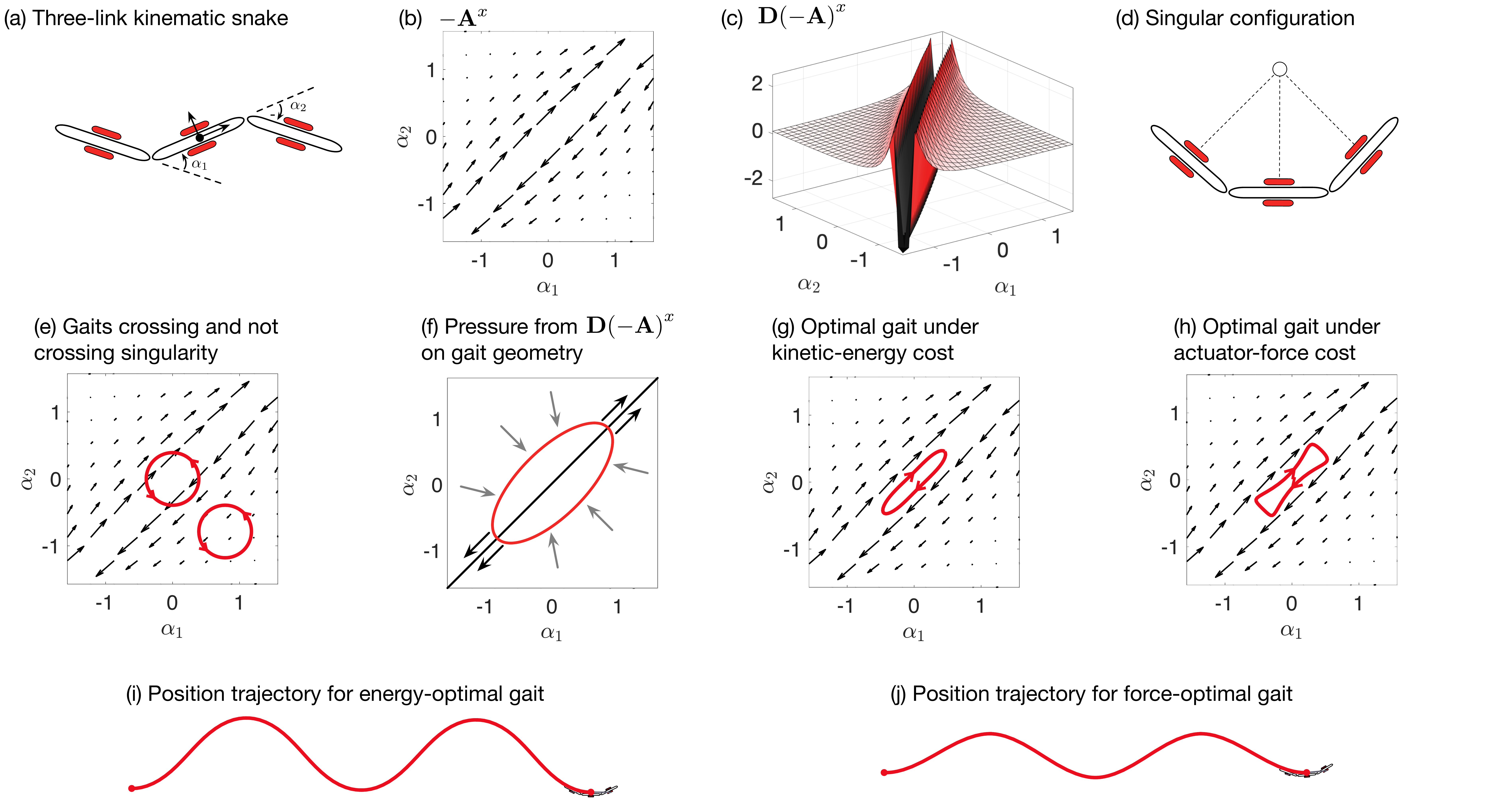}
\caption{The three-link kinematic snake (a) has a set of passive wheels on each link, preventing the link from moving laterally. The resulting local connection (b) and curvature of the local connection (c) have discontinuities at the $\alpha_{1}=\alpha_{2}$ line, corresponding to configurations in which the lines of constraint pass through a common center (d). Gaits can only cross this singularity orthogonal to it; as illustrated in (e), positively oriented gaits that cross the singularity go ``against the flow" of the field, whereas gaits that do not cross the singularity go ``with the flow". For singularity-crossing gaits, the optimization pressure from the CCF is to stretch along the discontinuity (enclosing the singular ``valley" in the CCF) while pinching inward everywhere else (excluding the opposite-sign regions on either side of the valley), as illustrated in (f). The specific shapes of the optimal gaits under kinetic-energy (g) and actuator-force (h) measures of effort are driven by their ``membrane-like" and ``beam-like" dynamics; in particular the gait in (h) pulls away from the singularity to avoid the high cost of accelerating in its vicinity. As illustrated in (i) and (j), the optimal gaits trace out sinuous paths through the world.}
\label{fig:kinematicsnake}
\end{center}
\end{figure*}

In the text above, we extracted both the locomotion model (encoded in $\mixedconn$) and the cost model (encoded in $\metric_{\base}$) from the kinetic energy metric $\metric$ over the full (shape and position) configuration space. There also exist systems %
for which the inertial costs of motion we present in this paper make good cost models, but whose locomotion models are not generated by applying conservation of momentum to the system. A canonical example of such systems is the three-link \emph{kinematic snake}~\cite{ostrowski_98,Shammas:2007,Hatton:2011IJRR,Hatton:2015EPJ,dear2020locomotion}, which, as illustrated
in Fig.~\ref{fig:kinematicsnake}(a), is a three-link chain with a passive wheelset attached to each link. These wheels act as direct nonholonomic constraints on the system motion (prohibiting system velocities in which any wheelset slides laterally) and together these constraints define a local connection $\mixedconn$ for the system of the same form as in~\eqref{eq:kinrecon}, whose structure is illustrated in Fig.~\ref{fig:kinematicsnake}(b).

Although this local connection is not itself derived from the system's inertial dynamics, we can use it to pull back the full inertia matrix into a reduced inertia matrix via the same process as in~\eqref{eq:kineticenergyreduced}. This reduced inertia matrix then provides a metric for evaluating the kinetic energy that it takes to move with a given shape velocity in the presence of the wheel constraints, and how much covariant acceleration and actuator force is required to follow a given shape trajectory.

The constraint curvature for the kinematic snake can be analyzed just as it was for the other two-joint systems, except on the $\alpha_{1}=\alpha_{2}$ line, where the local connection becomes singular, as illustrated in Fig.~\ref{fig:kinematicsnake}(c).\footnote{Singularities also appear at $\alpha_{1} = \pm \pi$ and $\alpha_{2} = \pm \pi$, but these self-colliding configurations are outside the scope of our discussion here.} Physically, these singularities correspond to the system forming a shape in which the lines normal to the wheel constraints meet at a single point, as illustrated in Fig.~\ref{fig:kinematicsnake}(d). This alignment means that the system can rotate around the convergence point (move in position space without changing shape) and can only cross the $\alpha_{1}=\alpha_{2}$ line at right angles to it.

Mathematically, the singularity appears in the system dynamics as the local connection $\mixedconn$ becoming asymptotically parallel to the singularity, with magnitude increasing towards infinity at the singularity, and with opposite sign on either side of the singularity. In the constraint curvature $\covextd{(-\mixedconn)}$, which is calculated as a derivative of $-\mixedconn$, the singularity appears in the $x$ component as an asymptotic rise in magnitude with the \emph{same} sign on both sides of the singularity, but with an opposite-sign ``delta function" at the singularity.

This delta-function structure %
corresponds to the ``well" at the center of the perfect-fluid system's $\covextd{(-\mixedconn)}^{x}$ CCF in Fig.~\ref{fig:CCFs}. %
As the ratio of lateral added mass to longitudinal and rotational added mass increases, this well becomes increasingly narrow; the kinematic snake represents a limit-case in which the lateral added mass becomes infinitely large, providing infinite resistance to the lateral motion of the links.

The integral of the constraint curvature inside a gait for the kinematic snake includes both the surface integral used in the earlier examples in this paper and a line integral of delta functions over the segment of the singularity that lies within the gait. Including this segment (which encodes the discrete change in $\mixedconn$ across the singularity) in the integral is analogous to treating corners in a line or creases in a surface as discrete equivalents of their geometric curvature when integrating to find net change in angle along a line or the net solid angle subtended by a surface.

Note that as illustrated in Fig.~\ref{fig:kinematicsnake}(e) the net line integral on $-\mixedconn$ for a gait that crosses the singularity is negative for counter-clockwise loops (and corresponds to the change in direction of the field across the singularity), whereas the integral for non-crossing gaits is positive for counter-clockwise loops (and corresponds to the fields' change in magnitude within each region). This difference highlights the fundamental importance of including the singularity's effect on $\covextd{(-\mixedconn)}$ in any constraint curvature analysis involving gaits that cross the $\alpha_{1}=\alpha_{2}$ line. In particular, it is important to recognize that although the limit of $\covextd{(-\mixedconn)^{x}}$ is \emph{positive} as the singularity is approached from both sides, the singularity itself is sufficiently \emph{negative} so as to make any integral containing it also negative; simply ``bridging over" the singularity by taking the value of $\covextd{(-\mixedconn)}$ as the mean of its limit-value on either side would result in an incorrect prediction of the direction of net displacement.

As illustrated in Fig.~\ref{fig:kinematicsnake}(f), the line integral applies pressure on gait curves to expand along the $\alpha_{1}=\alpha_{2}$ line, whereas the opposite-sign constraint curvature elsewhere in the shape space pushes the gait inwards. The optimal gaits for this system under the kinetic-energy and actuator-force metrics illustrate the effects of these pressures: The gait curves stay very close to the singularity, minimizing the amount of positive constraint curvature they enclose, which would otherwise cancel out some of the benefit of enclosing the singularity (On the plot of $-\mixedconn^{x}$, we see that staying close to the singularity means the gait is staying in a region where the magnitude of $-\mixedconn^{x}$ is large).

The optimal gait under the kinetic-energy measure of effort, illustrated in Fig.~\ref{fig:kinematicsnake}(g), has an oblong shape, and is prevented from straying too close to the singularity by the large amount of kinetic energy that changing shape in near-singular configurations entails. The optimal gait under the force-based measure of energy, illustrated in Fig.~\ref{fig:kinematicsnake}(g), bulges out at the ends to reduce the system's covariant acceleration while changing direction. The growth along the singularity of both of these gaits is ultimately constrained by the need to prevent their world trajectories, illustrated in Fig.~\ref{fig:kinematicsnake}(i) and (j) from becoming too sinuous, wasting motion in the lateral directions.

\section{Conclusions}

In this paper, we have presented a set of geometric principles defining the shape of optimal trajectories for isolated and perfect-fluid locomoting systems under ``least-action" and ``least-squared-acceleration" objective functions.

A key feature of this geometric framework is that it provides a fair comparison between ``small" gaits that can be repeated at high frequency and ``large" gaits that produce more displacement, but also require more time to execute each cycle: Fixing the average instantaneous cost of motion assigns each gait path a best-case-time-to-execute ``opportunity cost" related to its size; dividing the gait's induced displacement by this time provides a measure of gait effectiveness that is independent of any artifacts of a ``displacement per cycle" analysis.

In comparing the covariant acceleration cost function with the torque cost function, this framework also highlights the difference between optimizing for properties that are fundamental properties of the mechanism and those that also depend on how the actuators are attached to the mechanism.

Our extension of the inertial cost functions to nonholonomically-constrained systems such as the kinematic snake highlights the independence of the inertial cost formulation from the inertial dynamics formulation. It additionally demonstrates that the cost formulations continue to provide geometric insight in the presence of singularities in the dynamics.

Together with our previous development of such rules for systems with viscosity-dominated physics, we feel that this work provides a ``complete picture" of the planar locomotion of these kinematic locomoting systems, relating the path and pacing of their optimal gaits to the fundamental geometric structure of their physics.

Although we see this work as ``completing" one line of research, we do not see it as ``closing off" work in this area. Rather, we see it as establishing a ``well-furnished basecamp" from which to stage further investigations. Much research (some of it ours) has branched off from the same sources as we have drawn on for this work, and we see a place for our results here to provide either an explanation for observed results or as a point of comparison for how changing the physics of a system (e.g., by using more accurate models of fluid dynamics, introducing non-zero momentum conditions, or considering systems that are not described by ideal kinematic locomotion models) or introducing other cost functions changes a system's optimal motions. %

\appendices

\section{Energy-Optimal pacing} \label{app:optpacing}
In \S\ref{sec:kineticenergycost}, we use the quadratic nature of the kinetic-energy cost function to justify ``constant inertial speed" as a property of trajectories that minimize time to traverse a path at a given average kinetic energy. Taking the period of motion as $T$ and its inertial pathlength as $S$, the corresponding optimization can be stated formally as

\begin{align}
\text{minimize} \qquad  & T =  S/\bar{v} \label{eq:energycost}%
\\
\text{subject to constraint} \qquad & \frac{1}{T}\int_{0}^{T} v^{2}(t)\, dt = K,
\end{align}
where $v$ is the system's inertial speed, $\bar{v}$ is its time-averaged value, and $K$ is a constant.
Expanding this constraint in terms of a fluctuating velocity
\beq \label{eq:vsplit}
v(t) = \bar{v} + \tilde{v}(t),
\eeq
with the speed fluctuation constrained as
\beq \label{eq:fluconstraint}
\int_{0}^{T} \tilde{v}(t)\, dt = 0,
\eeq
gives
\beq \label{eq:resolvedfixedenergy}
\frac{1}{T} \bar{v}^{2} T + 0 + \int_{0}^{T} \tilde{v}^{2}(t)\, dt = K,
\eeq
and thus
\beq
\bar{v}^{2} = K-\int_{0}^{T} \tilde{v}^{2}(t).
\eeq
Because all terms in~\eqref{eq:resolvedfixedenergy} are positive, $K$ is necessarily greater than either of the other terms. The squared mean speed is thus maximized (and hence the time-to-execute cost in~\eqref{eq:energycost} is minimized) when $\tilde{v}(t) = 0$ such that the system moves with constant inertial speed.

\section{Acceleration-optimal pacing} \label{app:optpacingaccel}
In \S\ref{sec:accelcost}, we use the quadratic nature of the kinetic-energy cost function to justify ``slowing down in curved sections" as a property of trajectories that minimize time to traverse a path at a given average covariant acceleration.

To formally demonstrate this property, we can first use the unit-time/unit effort quartic proportionality from~\eqref{eq:time_period_constraint} to convert the constraint on the gaits we consider from fixed-effort to fixed-time. We can then convert the cost function from an integral over time to an integral over the path as
\begin{align}
\inertialcost^{a} &=\int_{0}^{T} \bigl( \dot{v}^{2}(t) + \kappa^{2}(t)v^{4}(t) \bigr)\, dt \label{eq:accelcost}\\
&=\int_{0}^{S} \tfrac{\partial t}{\partial s} \bigl( \dot{v}^{2}(s) + \kappa^{2}(s)v^{4}(s) \bigr)\, ds \\
&=\int_{0}^{S}  \Bigl( (v')^{2} v(s) + \kappa^{2}(s)v^{3}(s) \Bigr)\, ds,
\end{align}
where the final step is based on the identities $\tfrac{\partial t}{\partial s} = 1/v(s)$ and $\dot{v}(s)  =\tfrac{\partial{v(s)}}{\partial{s}} v(s) = v'(s) v(s) $.

The derivative of this cost with respect to a variation $\delta$ in the velocity $v(s)$ is
\begin{align}
\frac{\partial \inertialcost^{a}}{\partial \delta} &= \int_{0}^{S} 2 v' (\tfrac{\partial}{\partial \delta} v' )+ (v')^{2} (\tfrac{\partial}{\partial \delta} v ) + \kappa^{2}(3v^{2})(\tfrac{\partial}{\partial \delta} v ) \, ds
\\
&= \int_{0}^{S}  2 v' (\tfrac{\partial}{\partial \delta} v' ) + \left( (v')^{2} + (3\kappa^{2}v^{2})\right)  (\tfrac{\partial}{\partial \delta} v ). \label{eq:accelcostvariation}
\end{align}

We demonstrate that moving with constant inertial speed is not an equilibrium trajectory for acceleration-cost with respect to velocity variations that move the system slower in curved portions of the path by noting that for a constant-speed trajectory $v(s)=v_{0}$, the accompanying $v'(s) = 0$ condition means that the derivative in~\eqref{eq:accelcostvariation} does not depend on $(\tfrac{\partial}{\partial \delta} v' )$,
\beq
\left.\frac{\partial \inertialcost^{a}}{\partial \delta}\right|_{v(s)=v_{0}} = 3v_{0}^{2} \int_{0}^{S} \kappa^{2}(s) (\tfrac{\partial}{\partial \delta} v(s) )\, ds, \label{eq:costdeltaderivative}
\eeq
meaning that there is locally no cost for variations introducing $v'$ speed-spatial-acceleration, but that there is a benefit to decreasing speed in more-curved sections of the path. To see this benefit directly, we can introduce a velocity variation
\beq
v(s,\delta) = \bigl({1-\delta\kappa^{2}(s)}\bigr)\left(\frac{1}{T}\int_{0}^{S} \frac{1}{1-\delta\kappa^{2}(s)}\, ds \right),
\eeq
in which the left-hand term reduces the speed in proportion to the squared curvature of the path, and the right-hand term rescales the resulting velocities to keep the mean velocity constant, and which has $v'(s)=0$ when $\delta=0$. The derivative of the velocity with respect to the variations at $\delta=0$ resolves to
\beq \label{eq:velocityvariationderivative}
\left.\frac{\partial v(s,\delta)}{\partial \delta}\right|_{\delta=0} = v_{0}\bigl(\kappa^{2}_{\text{av}}-\kappa^{2}(s)\bigr),
\eeq
(where $v_{0}=S/T$), i.e. the system slows down in regions that have a greater-than-average squared curvature, while speeding up in regions with smaller-than-average squared curvature. Inserting this velocity derivative into the cost derivative from~\eqref{eq:costdeltaderivative} gives the cost derivative with respect to this variation as
\beq
\left.\frac{\partial \inertialcost^{\tau}}{\partial \delta}\right|_{\delta=0} = 3v_{0}^{2} \int_{0}^{S} \kappa^{2}(s) \bigl(\kappa^{2}_{\text{av}}-\kappa^{2}(s)\bigr)\, ds.
\eeq
Because the $\kappa^{2}(s)$ term scales negative regions of $\bigl(\kappa^{2}_{\text{av}}-\kappa^{2}(s)\bigr)$ by a greater magnitude than it scales positive regions, the integral is guaranteed to be non-positive, and is strictly negative outside of the constant-squared-curvature case where $\kappa^{2}(s) = \kappa^{2}_{\text{av}}$.

\section{Gait parameterization:}
For several of our calculations (e.g., the gradient calculation below in Appendix~\ref{app:actuatorforcetimegradient}, it is useful to parameterize the systems' gaits via truncated Fourier series, such that the shape at time $t$ in the gait is calculated as

\begin{equation}
\base_i(t) = a_0 + \sum_{j=1}^{k} \left( a_j\cos{j\omega t} + b_j\sin{j\omega t} \right).
\label{eq:shape_position_fourier}
\end{equation}
The shape velocity and acceleration at time $t$ are then easily determined through differentiation of this expression:
\begin{equation}
\dot{\base}_i(t) = \sum_{j=1}^{k} \left( -j\omega a_j\sin{(j\omega t)} + j\omega b_j\cos{(j\omega t)} \right),
\label{eq:shape_velocity_fourier}
\end{equation}
and
\begin{equation}
\ddot{\base}_i(t) = \sum_{j=1}^{k} \left( -(j\omega)^2 a_j\cos{(j\omega t)} - (j\omega)^2 b_j\sin{(j\omega t)} \right).
\label{eq:shape_acceleration_fourier}
\end{equation}

\section{Gradient of the Time Period for Actuator-Force Cost} \label{app:actuatorforcetimegradient}

The second term of~\eqref{eq:gradient1} takes $\nabla_p T$ as a measure of how moving points in the gait path gait influences the acceleration cost of executing the gait. Taking the gradient of the actuator-force-constrained period $T$ from~\eqref{eq:time_period_constraint} with respect to the parametrization $p$ gives the gradient of the gait's period in terms of the unit-period cost of the gait at with the current parameters and the gradient of the unit period cost with respect to the parameters,
\begin{align}
\nabla_{p} T &= \nabla_{p} \left( \int_{0}^{1} \tau_{1}^{2} dt \right)^{1/4} \\
&= \frac{1}{4}\left( \int_{0}^{1} \tau_{1}^{2} dt \right)^{-3/4} \left( \nabla_{p} \left( \int_{0}^{1} \tau_{1}^{2} dt \right) \right),
\end{align}
which, by replacing the first integral expression with $T$ via~\eqref{eq:time_period_constraint}, simplifies as
\begin{equation}
\nabla_{p} T = \frac{1}{4T^3} \int_{0}^{1} \left( 2 \tau_1 \nabla_{p} \tau_{1} \right) dt.
\end{equation}

Because the quantities $T$ and $\tau_1$ are known at each step in the optimization, computing the gradient $\nabla_{p} T$ requires only further calculation of  $\nabla_{p}\tau_{1}$. Applying the gradient $\nabla_{p}$ to each of the terms in~\eqref{eq:EL} and~\eqref{eq:ELC} via the chain rule gives
\begin{subequations}
\label{eq:gradient_of_cost}
\begin{align}
\nabla_{p} \tau_{1} &= \left( \nabla_{p} \metric_\base(\base) \right)\ddot{\base} \\
&\phantom{=}+ \metric_\base(\base) \left( \nabla_{p} \ddot{\base} \right) \\
&\phantom{=}+ \sum_i \left( \left( \nabla_{p} \frac{\partial \metric_\base(\base)}{\partial \base_i} \right) \dot{\base}_i  + \frac{\partial \metric_\base(\base)}{\partial \base_i} \left( \nabla_{p} \dot{\base}_i \right) \right) \dot{\base} \\
&\phantom{=}+ \sum_i \left( \frac{\partial \metric_\base(\base)}{\partial \base_i}\dot{\base}_i \right) \left( \nabla_{p} \dot{\base} \right) \\
&\phantom{=}- \frac{1}{2} \left( \left( \nabla_p \dot{\base} \right)^T \frac{\partial \metric_\base(\base)}{\partial \base} \dot{\base}  + \dot{\base}^T \left( \nabla_p \frac{\partial \metric_r(\base)}{\partial \base} \right) \dot{\base} \right) \\
&\phantom{=}-\frac{1}{2} \dot{\base}^T \frac{\partial \metric_r(\base)}{\partial \base} \left( \nabla_{p} \dot{\base} \right),
\end{align}
\end{subequations}
from which $\nabla_{p} \tau_1$ can be obtained by constraining the gait defined by $\base$ to be completed in unit time.

This expression may be further rearranged such that the gradient operation with respect to the parameters is applied only to the path variables by expanding $\nabla_{p} \metric_r(\base)$ and $\nabla_{p} \frac{\partial \metric_r(\base)}{\partial \base}$ using the chain rule as
\begin{align}
\nabla_{p} \metric_r(\base) &= \left( \frac{\partial \metric_r(\base)}{\partial \base} \right) \left( \nabla_{p} \base \right)\\ &= \sum_i \left( \frac{\partial \metric_r(\base)}{\partial \base_i} \nabla_{p} \base_i \right)
\label{eq:gradient_of_mass_matrix}
\end{align}
and
\begin{align}
\nabla_{p} \frac{\partial \metric_r(\base)}{\partial \base} &= \left( \frac{\partial^2 \metric_r(\base)}{\partial \base^2} \right) \left( \nabla_{p} \base \right) \\
&= \begin{bmatrix}
\sum_{j} \left( \frac{\partial^2 \metric_r(\base)}{\partial \base_1 \partial \base_j} \nabla_{p} \base_j \right) \\
\vdots \\
\sum_{j} \left( \frac{\partial^2 \metric_r(\base)}{\partial \base_n \partial \base_j} \nabla_{p} \base_j \right) \\
\end{bmatrix}.
\label{eq:gradient_of_partial_mass}
\end{align}

\subsection*{Fourier implementation of the shape velocity and acceleration:} Because substituting ~\eqref{eq:gradient_of_mass_matrix} and~\eqref{eq:gradient_of_partial_mass} into~\eqref{eq:gradient_of_cost} results in an expression that only requires derivatives of $M$ with respect to $\base$ and $\base$ with respect to $p$, we implement the gradient calculation via a truncated Fourier series parameterization for the gaits. Each term of these gradients is composed of a partial derivative $\frac{\partial}{\partial f}$, where $f$ is the a given Fourier coefficient $a_j$ or $b_j$.

The terms for the gradient of shape $\nabla_{p} r$ are then calculated by differentiating \eqref{eq:shape_position_fourier} with respect to $a_k$ and $b_k$, resulting expressions
\begin{equation}
\frac{\partial \base_{i}(t)}{\partial a_k} =%
\cos{(k\omega t)},%
\end{equation}
and
\begin{equation}
\frac{\partial \base_{i}(t)}{\partial b_k} = \sin{(k\omega t)}.
\end{equation}
Similarly, differentiating~\eqref{eq:shape_velocity_fourier} provides the terms for $\nabla_{p} \dot{\base}$:
\begin{equation}
\frac{\partial \dot{\base_{i}}(t)}{\partial a_k} = %
-(k\omega)\sin{(k\omega t)},%
\end{equation}
and
\begin{equation}
\frac{\partial \dot{\base}(t)}{\partial b_j} = (j\omega)\cos{(j\omega t)}.
\end{equation}
Finally, differentiating~\eqref{eq:shape_acceleration_fourier} provides the terms for $\nabla_{p} \ddot{\base}$:
\begin{equation}
\frac{\partial \ddot{\base_{i}}(t)}{\partial a_k} = %
-(k\omega)^2\cos{(k\omega t)}, %
\end{equation}
with
\begin{equation}
\frac{\partial \ddot{\base_{i}}(t)}{\partial b_k} = -(k\omega)^2\sin{(k\omega t)},
\end{equation}
(where for each shape variable, the index $k$ is scoped to the set of Fourier coefficients corresponding to that shape variable).

\section{Actuator-force Metric for Viscous Systems}

In~\cite{Hatton:2017TRO:Cartography,Ramasamy:2019aa}, we measured the cost of a gait for a viscous system as the energy dissipated through viscous friction between the swimmer and its environment. This cost translated into geometric form as the pathlength of the gait under a Riemannian metric $M_{\base}$ constructed by reducing the total system drag matrix via the local connection. This drag cost is analogous to the covariant-acceleration cost for the inertial systems, in that it considers only the force acting on each particle of the system at that particle's location, but does not account for the leverage the actuators have on the body. The actuator-force cost for the viscous system is the pathlength under the \emph{square} of the reduced drag matrix,
\beq
\text{cost} = \int_{\gait} \overbrace{\basedot^{T} M}^{\phantom{^{T}}\tau^{T}}\overbrace{ M \basedot\vphantom{^{T}}}^{\tau}.
\eeq
The optimal gaits under this squared metric are qualitatively the same as those under the original metric, and are slightly shorter (when viewed in the parameter space) along the even axes of the shape space (where the true metric geometry is ``long" relative to the coordinate geometry).
\bibliographystyle{IEEEtran}

\end{document}